\RequirePackage{letltxmacro}
\LetLtxMacro{\LaTeXtextbf}{\textbf}
\documentclass{ieeeaccess}
\LetLtxMacro{\textbf}{\LaTeXtextbf}

\IEEEoverridecommandlockouts

\usepackage{cite}
\usepackage{amsmath,amssymb,amsfonts}
\usepackage{algorithmic}
\usepackage{graphicx}
\usepackage{textcomp}
\usepackage{xcolor}

\newcommand\hl[1]{#1}

\usepackage[utf8]{inputenc}

\def\BibTeX{{\rm B\kern-.05em{\sc i\kern-.025em b}\kern-.08em
    T\kern-.1667em\lower.7ex\hbox{E}\kern-.125emX}}

\def\thevol{}
\def\theyear{2022}

\usepackage{multirow}
\usepackage[inline]{enumitem}
\usepackage[ruled]{algorithm2e}
\usepackage{hyperref}

\begin{document}

\history{Date of publication xxxx 00, 0000, date of current version xxxx 00, 0000.}
\doi{10.1109/ACCESS.2022.3147312}

\title{Automatically generating models of IT systems}

\author{IVAN KOVAČEVIĆ\authorrefmark{1},
STJEPAN GROŠ\authorrefmark{2}, and ANTE DEREK\authorrefmark{3}}
\address[1]{
University of Zagreb Faculty of Electrical Engineering and Computing,
Zagreb, Croatia (e-mail: ivan.kovacevic@fer.hr)}
\address[2]{
University of Zagreb Faculty of Electrical Engineering and Computing,
Zagreb, Croatia (e-mail: stjepan.gros@fer.hr)}
\address[3]{
University of Zagreb Faculty of Electrical Engineering and Computing,
Zagreb, Croatia (e-mail: ante.derek@fer.hr)}

\markboth
{Kovačević \headeretal: Automatically generating models of IT systems}
{Kovačević \headeretal: Automatically generating models of IT systems}

\corresp{Corresponding author: Ivan Kovačević (e-mail: ivan.kovacevic@fer.hr).}

\tfootnote{This work was supported in part by 
the research and development project Cyber Conflict Simulator, co-financed by the EU under Grant KK.01.2.1.01.0054.}

\begin{abstract}
Information technology system (ITS), informally, consists of hardware and software infrastructure (e.g., workstations, servers, laptops, installed software packages, databases, LANs, firewalls, etc.), along with physical and logical connections and inter-dependencies between various items. Nowadays, every company owns and operates an ITS, but detailed information about the system is rarely publicly available. However, there are many situations where the availability of such data would be beneficial. For example, cyber ranges need descriptions of complex realistic IT systems in order to provide an effective training and education platform. Furthermore, various algorithms in cybersecurity, in particular attack tree generation, need to be validated on realistic models of IT systems. In this paper, we describe a system we call \textit{the Generator} that, based on the high-level requirements such as the number of employees and the business area the target company belongs to, generates a model of an ITS that satisfies the given requirements. We  put special emphasis on the following two criteria: the generated ITS models a large amount of details, and ideally resembles a real system. \hl{Our survey of related literature found no sufficiently similar prior works, so we believe that this is the first attempt of building} something like this. We created a proof-of-concept implementation of the Generator, validated it by generating ITS models for a simplified fictional financial institution, and analyzed the Generators performance with respect to the problem size. \hl{The research was done in an iterative manner, with coauthors continuously providing feedback on intermediate results}. The conducted experiments show that our approach is feasible. In the future, we intend to extend this prototype to allow probabilistic generation of IT systems when only a subset of parameters is explicitly defined, and further \hl{develop and} validate our approach with the help of domain experts.
\end{abstract}




\titlepgskip=-15pt

\maketitle


\section{Introduction}
\label{sec:introduction}

Practice is key in the learning process and Cyber Defense is no exception. Knowing your information system well, knowing what would happen in the event of an attack, testing defenses and practicing with a simulated adversary on the network are critical to a successful defense. But practicing on production systems is prohibited for availability, integrity, and confidentiality reasons, and rarely can anyone afford duplicate systems just for testing purposes. For this reason, cyber ranges \cite{yamin2020,forrester,ukwandu2020,russo2020} have emerged as a viable alternative. Cyber ranges are essentially virtualized environments that mimic real-world environments and also contain additions that make practicing and learning more efficient. They are a great tool that has been used in cybersecurity for some time, and as time goes on, they will be used even more \cite{forrester}. Cyber ranges also help in cases where new security professionals are being trained, as they provide a near-realistic environment in which they can test and improve their skills. Finally, cyber ranges can also be used to test different security mechanisms and how they behave in the event of an attack.

All of this requires that an organisation's information technology system (IT system) --- whether real or imagined --- be implemented in the cyber range. And this can be done in two ways. The first approach is to take a snapshot of an existing IT system and model it in a cyber range. This is relatively easy to do because various network discovery and scanning tools can be used. These tools are able to create machine-readable descriptions of what is detected, which can then be translated, again using other tools, into a virtual network description that is implemented in the cyber range. However, existing IT systems are not always readily available. While there are many published network datasets in various domains~\cite{snapnets, nr}, we are not aware of any dataset that contains descriptions of complex IT systems.


But building an IT system from scratch, even of a moderate size, is very resource intensive and requires a great deal of knowledge. Real information systems and the IT systems that support them are very complex and contain a lot of detail. An additional problem is that IT systems and the information systems that support them can be created in an almost infinite number of ways. However, not all of them are valid or even useful, as there are some rules and blueprints that are followed when building an IT system. Moreover, IT systems also grow organically, meaning that starting from scratch is not always an option, more often information systems and IT systems that support them grow with the growth of the business they support. Because of this, some configurations pop up in practice that would not in IT systems designed from scratch. All of this makes creating complex realistic environments for practice in cyber ranges very difficult, with the result that when IT systems are created within a cyber range a very limited set of configurations is used, which affects the learning process.


There are several research areas in computer security that can directly benefit from being able to easily generate detailed models of IT systems of arbitrary size. The most obvious application is the generation of cyber ranges, where the existing systems (e.g. \cite{schreuders2017, eckroth2019}) could be used to build ready-to-use ranges based on the models of IT systems, such as those generated by the system we propose in this paper. Furthermore, models of IT systems can be used to test and evaluate procedures for generating and analyzing attack graphs \cite{sheyner2002, Xinming2006}. Automated security risk analysis would also benefit from this work. For example, \cite{singhal2017} and \cite{homer2009} develop methods for enterprise networks, but then they test it on a few very simple networks --- the models of IT systems generated by our approach are much more similar to the enterprise networks they wish to target. Finally, machine learning approaches to evaluating attacks and defenses (e.g., \cite{ghanem2020, bland2020} just to mention a few) can benefit from realistic models of IT systems both for training and evaluation, especially in the case of reinforcement learning where the results depend of availability of various models of IT systems representative of real-world systems.

In this paper we present a first step towards solving the problem of a finite set of different IT systems. To this end, we describe a prototype \hl{expert} system we have developed that, given a set of high level requirements, produces a description of an IT system. In summary, our contributions are as follows:

\begin{itemize}
    \item Proposal of a concept of an IT system generator.
    \item Rules that regulate how certain objects are instantiated and linked.
    \item Proof-of-concept implementation of IT system generator.
    \item Dataset containing sample generated models of ITSs \cite{its-dataset}.
\end{itemize}

We also evaluated the performance of the proof-of-concept implementation for a different problem sizes. The results show that CPU consumption and memory usage are both acceptable for intended ranges of inputs.

The paper is structured as follows. First, in Section \ref{sec:background} we define IT system and also give a quick overview of some technologies used for the implementation of a generator. Then, in Section \ref{sec:principles} we give a high-level description of the system we want to build and its principles. In Section \ref{sec:generator} we describe the architecture and implementation of the proposed generator system, followed by its validation in Section \ref{sec:validation}. Section \ref{sec:performance} describes our performance experiments and measured behavior of the prototype system for the given inputs. In Section \ref{sec:discussion} we analyze the obtained results and discuss possibilities and use cases of our approach. We also discuss the limitations of the proposed system, and features that still need to be implemented. The subsequent Section \ref{sec:relatedwork} provides a review of related work. Finally, the paper finishes with conclusions and future work in Section \ref{sec:conclusion}, followed by the list of references.

\section{Background}
\label{sec:background}

In this section, we define the term information technology system model (or a model of an information technology system) and also give a brief description of two key technologies, SMT solvers and integer linear programming tools, that we used to implement the generator. 

In general, \textit{information technology} (IT) is defined as a set of interrelated technologies for information processing. \textit{Technologies} mean software, hardware and communication technologies \cite{gartner}. This corresponds exactly to what we want to achieve, i.e., we want to develop a system that, given certain input requirements, generates a description of an information technology system (ITS) consisting of workstations and servers, software installed on them, printers, a properly segmented computer network connecting the computers, and specialized security devices such as firewalls, \textit{\hl{Intrusion Detection Systems}} (IDS), \textit{\hl{Intrusion Prevention Systems}} (IPS), and so on. A \textit{model of an information technology system} is a simplified version of a real IT system in which details that are not important are omitted. For example, since we are interested in the application of IT systems in security, we do not model elements of IT systems that are irrelevant in the security application. Examples of such software include various types of load balancers and failover configurations that are not important from a security perspective (unless they themselves have vulnerabilities, which we ignore), so we do not attempt to model them. In the following text, we will use the term \textit{information technology system (ITS)} instead of \textit{model of information technology system (ITSM)} when there is no ambiguity about whether we are talking about a model generated by our system or a IT system as used in a particular environment.

\emph{Satisfiability Modulo Theory} (SMT) solving is a form of \emph{constraint solving}, in which constrains are modeled as logical formulas. More precisely, SMT problem consists of first order logical formulas interpreted in a \emph{background theory} such as booleans, bitvectors, integers (usually of finite size) etc. SMT solvers attempt to solve the constraints by producing a instantiation of values for variables that satisfies all the formulas. We use constraints to specify policies that the network infrastructure needs to satisfy. Our implementation uses the Z3 Theorem Prover~\cite{z3} from Microsoft Research that constructs network segments that satisfy the constraints.

\emph{Integer linear programming} (ILP) is an optimization problem where both the cost function and the constrains are linear, and the variables are restricted to integer values. We use ILP when selecting software components to be installed in the ITS --- the linear constraints implement the requirements (e.g., the combination of software must support required user services) and the cost function is the total cost of software licenses and maintenance. Our implementation uses the \hl{software package for convex optimization called} \textit{CVXOPT}~\cite{cvxopt}.

\section{Basic principles and requirements}
\label{sec:principles}

In this section, we provide a high-level overview of what \textit{we want to accomplish}, i.e., our long-term goals. We also lay out some basic principles and assumptions that will determine how we approach solving the problem. Based on these assumptions and principles, we have determined that templates, rules, and input parameters are required on the input, so we describe these in more detail in this section as well. In the following section, we describe what we implemented as a proof-of-concept and what we learned from it.

Our goal is to create a system that can generate models of real-world IT systems. To tell if a model of an IT system is real-world or not we envision a large set of artificially generated IT systems. Samples are again drawn randomly and presented independently to $N$ experts. Each expert has to identify which, if any, part of the IT system is not something that would appear in the real-world. If at least  50\% of experts for 50\% of samples can not identify anything differently from the real world IT systems, then the process of generating models of IT systems generates real-world IT systems. In this work we did not perform this test on a large scale, the only experts that were evaluating generated models of IT systems were the authors. We also strived to get 100\% of samples to have no difference between real-world systems and the ones artificially generated.

The way IT systems are created in businesses and institutions does not follow from any first principles or axioms, but rather is the result of experience, best practices, various restrictions (e.g., finances), personal preferences, and many other influences. Moreover, complex IT systems are rarely created from scratch in one step. The much more common case is that they grow organically as a company's information system they support grows. As technology evolves and organizations adopt new technologies, they are integrated into their IT systems not in an ideal way, but in the best possible way that the specific IT system allows and supports. It follows that in order to build models of IT systems in companies, we need to model these processes somehow. In other words, this means that we need to model the experts' thinking and their mental decision-making processes.

\begin{figure}
    \centerline{\includegraphics[width=8.5cm]{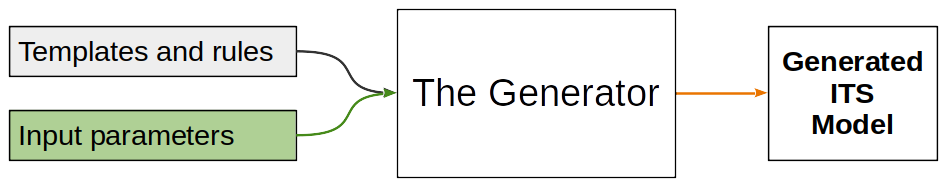}}
    \caption{Inputs and outputs of the Generator}
    \label{fig:generator}
\end{figure}

There are two classes of systems used to model human behavior. The first class is based on machine learning, and specifically deep learning. The problem with this approach is that it requires a huge amount of data, which is not available in our case. The other class is expert systems \cite{grosan2011}, which encode the knowledge of experts into a set of rules. The rules are then applied to imitate an expert. This is the approach we have taken. To consider blueprints used in building IT systems, we have chosen to use \textit{templates}. The idea is that templates represent a generic part of an IT system that are instantiated and specialized by applying rules for a particular need. Finally, we have parameters that are necessary because not everything can be encoded in rules and templates. For example, some parameters are the number of employees and the vertical industry that the generated IT system should support, e.g. banking, manufacturing, software company. Based on these premises, we envision the system shown in Figure \ref{fig:generator}, which we will call \textit{the Generator}, that takes some input parameters and a set of rules and templates, and generates one or more models of IT systems at the output.

Based on the inputs, the Generator creates a model of IT system which satisfies given requirements. Anything else that is not specifically defined is arbitrarily selected from some distribution. For example, if the use of Active Directory (AD) \cite{wiki:activedirectory} was not specified, or more generally centralized authentication system, as a requirement in the input, the Generator generates one based on a probability distribution indicating the likelihood of there being AD as a function of the number of workstations in an IT system. This means that the Generator can generate a potentially infinite number of different models of IT systems, all obeying a given input.

Aforementioned templates are described in Section \ref{sec:component-templates}, parameters in Sections \ref{sec:component-parameters} and \ref{sec:rules}, and rules in Section \ref{sec:rules}.


\subsection{Templates}
\label{sec:component-templates}

Templates describe software packages, employee roles, and organizational services. Software packages are described using attributes enumerated in Appendix \ref{appendix:template-attributes}. Among others, attributes describe the following information about software packages:

\begin{itemize}
  \item Local and network dependencies;
  \item Services provided to users on workstations (i.e. email access);
  \item An abstract measure of the amount of required hardware resources, that we refer to as \textit{hardware quota};
  \item Costs of licenses and maintenance.
\end{itemize}

Employee roles link employees to information about their local service requirements, which define what the software packages installed on their workstations must support. An example of such requirements would be a bank clerk who needs to be able to access financial services. Since such services are provided by financial client applications, the proposed system ensures that a financial client application is installed on the bank clerks's workstations.

Organizational services describe high-level services that the organization must provide to its external customers, such as Internet banking. Templates describing organizational services contain information describing what combinations of network services must be exposed to the Internet to support individual organizational services. For example, Internet banking, mentioned above, requires that an Internet banking Web application be installed and configured to accept connections from the Internet.

\subsection{Input parameters}
\label{sec:component-parameters}

In addition to the templates, which are intended to be generic and applicable to all supported types of organizations, the description of a concrete target organization is required to generate a ITS. We refer to this information as input parameters, and they include the following:

\begin{itemize}
  \item \textit{Employee role subgroups} (ERS), specifying subgroups of employees belonging to various employee roles that exist within the organization, and the number of employees in each subgroup;
  \item \textit{Data collections}, describing what data collections must be installed to support the organization;
  \item \textit{Provided external services}, listing external services that the organization must provide; and
  \item \textit{Network policies}, describing any custom rules that regulate what can, what must and what must not belong to the same network segment.
\end{itemize}

Data collections and network policies are further described in Subsection \ref{sec:rules}. The input parameters can be defined in any combination. For example, in one case only the number of employees could be specified, while in the other case the desired services and the number of customers could be specified. It could also be specified that the company is small, medium or large. In either case, the Generator should derive other parameters of the IT system it generates based on the given requirements. If this is not possible due to a lack of correlation between these parameters in the real world, the Generator is free to choose whatever it deems most appropriate. Rules can also be added or removed as needed, or as the Generator is improved.

\subsection{Rules}
\label{sec:rules}

One of our contributions are rules for dataset instantiation and network segmentation. Dataset instantiation is governed by dataset instantiation rules, which can be referenced by special input parameters called \textit{data collections}. Network segmentation is governed by \textit{network segmentation rules}, which can be divided into default rules embedded in the system and user-defined rules that can be selected and added in input parameters called network policies.

Data collections are templates that represent classes of data within an organization. They define which dataset instances are created and which employees and software they are associated with. \hl{Each data collection defines a \textit{protection level}, a numeric relative measure of importance which can be used to define policies.} Dataset instantiation rules support the following criteria for instantiation:

\begin{itemize}
  \item A dataset can be primarily stored either on a database, a server, or a client;
  \item A dataset can be installed for some organizational services, which requires it to appear on a server software that supports such organizational services;
  \item A dataset can be installed for some employee roles, which requires it to appear on client software installed for its member employees;
\end{itemize}

In addition, there can be several modes of linking dataset instances to other objects, for which additional requirements are supported:

\begin{itemize}
    \item \textit{Employee mode} refers to the mode of creating datasets with regard to employees, with three possible variants: \begin{enumerate*}
        \item each employee must have distinct dataset instances not shared with other employees,
        \item each ERS must have distinct dataset instances not shared with other ERSs, and
        \item each dataset instance must be shared between all employees;
    \end{enumerate*}
    \item \textit{Service mode} refers to the mode of creating datasets with regard to organizational services, with two possible variants: \begin{enumerate*}
        \item each organizational service must have distinct dataset instances not shared with other organizational services, and
        \item each dataset instance must be shared between all organizational services;
    \end{enumerate*}
    \item \textit{Software mode} refers to the mode of creating datasets with regard to software, with five possible variants: \begin{enumerate*}
        \item each database must have distinct dataset instances not shared with other databases,
        \item each server must have distinct dataset instances not shared with other servers,
        \item each client must have distinct dataset instances not shared with other clients,
        \item each software must have distinct dataset instances not shared with other software, and
        \item each dataset instance must be shared between all software in the organization.
    \end{enumerate*}
\end{itemize}

The connection between data collections and datasets can be illustrated by the example of emails. Emails are a data collection because they represent a class of data that applies to the entire organization. Their definition specifies that each employee who reads emails must have a corresponding email dataset instance that contains only their personal emails. Therefore, each final email dataset instance will be associated with an individual employee and their email client applications. In addition, each email dataset instance will be linked to the appropriate email server software and stored in the appropriate database. This linkage is very important during cyber exercises, for example, as it can help defenders narrow down the source of data leakage during a simulated data breach investigation.

Network segmentation rules specify requirements that govern the selection of network segments for dataset instances and software installations. Each rule can contain a set of queries that select objects to which the rule applies. Supported types of rules are the following:

\begin{itemize}
  \item \textit{RequireDistinctNetworksForSets(setA\_query, setB\_query)} requires that two sets of objects specified by queries \textit{setA\_query} and \textit{setB\_query} must never appear together inside a network segment;
  \item \textit{RequireSameNetworksForSets(setA\_query, setB\_query)} requires that two sets of objects specified by queries must always appear together in a network segment;
  \item \textit{RequireCommonNetworksForSets(setA\_query, setB\_query)} requires that objects selected by \textit{setA\_query} and objects selected by \textit{setB\_query} must have \textit{some} shared network segments, in which they appear together;
  \item \textit{RequireCollocatedLocalDependencies()} requires that software installations must \textit{at least once} appear in a network segment that contains their local requirements;
  \item \textit{LimitAllowedProtectionLevelRange(allowed\_difference)} requires that no pair of objects within a network segment has a difference between protection levels larger than the target limit;
\end{itemize}

The rule \textit{RequireCollocatedLocalDependencies} is relaxed, as denoted by "\textit{at least once}" above, to avoid conflicts during network segmentation. Suppose that servers A, B, and C are installed, and that C is a local dependency for both A and B. Furthermore, suppose that there are two dataset instances, D1 and D2, with D1 stored on A, and D2 stored on B and C, that are required to never appear together within a network segment. If the rule was not relaxed, the network could never be generated because A and B would both need to be located in all network segments that contain C, and that would break the requirement for splitting D1 and D2. With the relaxed rule, A and C can appear in one network segment containing D1, and B and C in another containing D2, making network segmentation possible.

During development, we defined default rules to avoid pointless network segment assignment and ensure basic network functionality common to most organizations. Defining a minimal set of rules is still ongoing work. These rules are the following:

\begin{itemize}
  \item Software packages that are only depencencies of other software packages must be inside network segments that contain some software packages that depend on them;
  \item Each software installation must have at least one common network segment with each related dataset instance, and dataset instances must not appear in network segments that contain no related software installations;
  \item Software installations that provide organizational services must at least once appear inside a network segment that accepts connections from the Internet;
  \item Dataset instances related to organizational services must at least once appear inside a network segment that accepts connections from the Internet;
\end{itemize}

\section{Generator Architecture and Implementation}
\label{sec:generator}

This section describes the proposed architecture of the Generator component and our implementation of that architecture.

\subsection{Architecture}
\label{sec:architecture}

\begin{figure*}
    \centering
    \includegraphics[width=17cm]{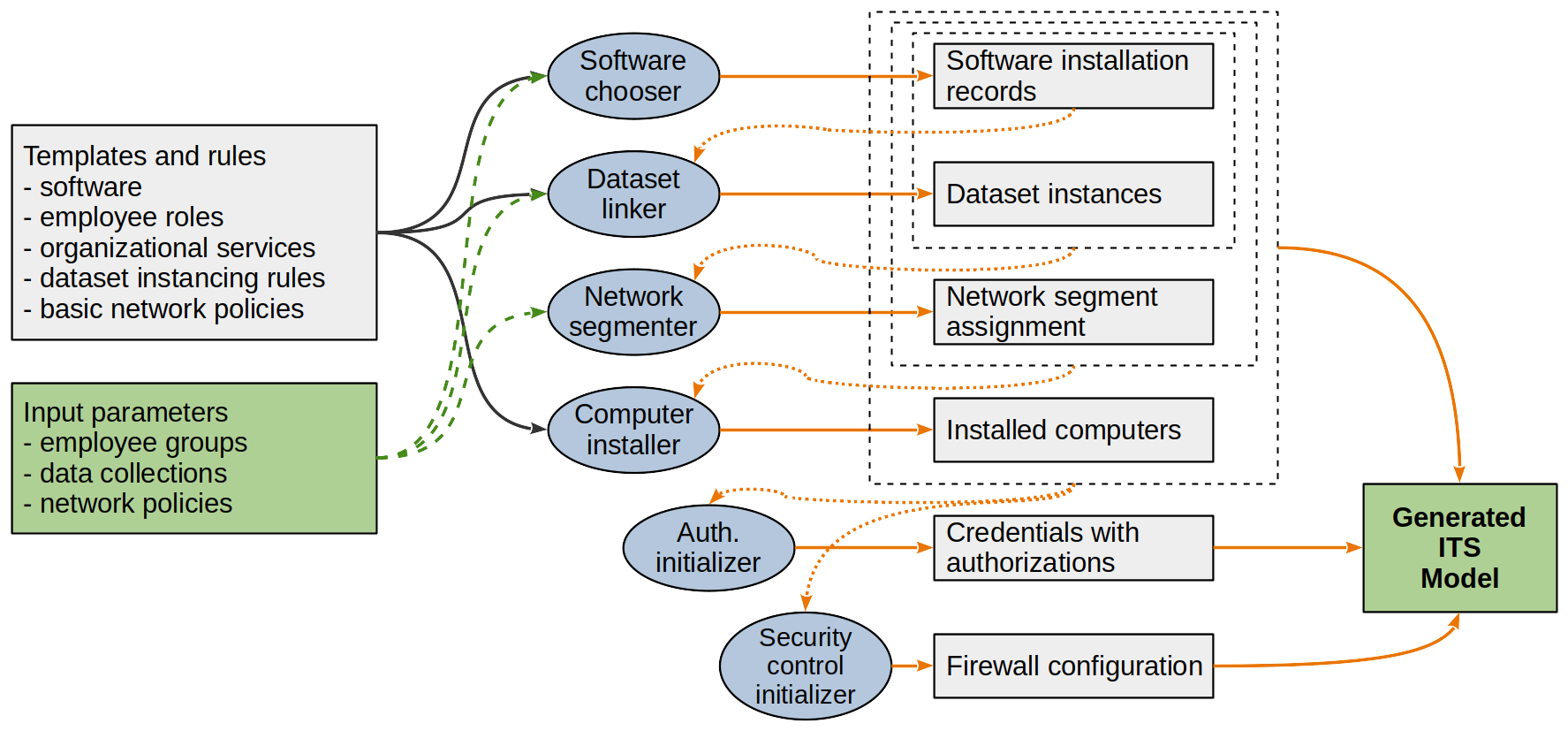}
    \caption{Architecture of the proposed method and system. Circles represent components of the system, rectangles represent data, and arrows represent the data flow. Dashed rectangles denote that multiple sets of data are used as inputs for a component.}
    \label{fig:architecture}
\end{figure*}

The proposed method and system can be broadly divided into six components corresponding to six phases of ITS generation, as shown in Figure \ref{fig:architecture}. These components, described in Subsections \ref{sec:component-software}-\ref{sec:component-security}, are the following:
\begin{enumerate*}
  \item \textit{Software chooser},
  \item \textit{Dataset linker},
  \item \textit{Network segmenter},
  \item \textit{Computer installer},
  \item \textit{Authentication initializer}, and
  \item \textit{Security control initializer}.
\end{enumerate*}

\subsubsection{Software chooser}
\label{sec:component-software}

The first task the system must perform is to select a set of software packages for installation. This was chosen as the first task because other objects such as datasets, computers, credentials, and firewall rules either reference or rely on existing software installations. For example, the number of server computers required cannot be determined without considering the hardware requirements of the software and associated datasets to be installed. Software is selected based on the following criteria:

\begin{enumerate}
  \item For each employee, software supporting all required user services must be installed on appropriate workstations;
  \item For each organizational service, a combination of software that supports all of it's required network services must be installed on a server computer;
  \item For each installed software, a software satisfying all its local requirements must be installed on the same computer;
  \item For each installed software, server software satisfying all of its network requirements must be installed on a server;
  \item The costs of software licenses and maintenance must be as low as possible, so that a minimal set of software satisfying all previous requirements is installed.
\end{enumerate}

It should be noted that requirements of employees and organizational services must allow for multiple variations of software packages. For example, if there are two email client applications and an employee requires email support, it must be possible to select any of the two email client applications for installation. The output of this component is a collection of software installation records that contain data from the appropriate software templates that are either associated with the employees for whom they are installed or designated as server software.

\subsubsection{Dataset linker}
\label{sec:component-datasets}

After the software installation records have been created, the instantiation and linking of the datasets can take place. All data collections are analyzed and instances are created according to the input parameters described in Subsection \ref{sec:rules}. If required, separate instances are created for different combinations of employees, organizational services, and software. The output of this phase is a collection of dataset instance records that reference linked employees, organizational services, and software installation records previously created by the software selection component.

\subsubsection{Network segmenter}
\label{sec:component-network}

After software installation records and dataset instance records are created, the network segmentation component attempts to create a configuration of network segments that conforms to the specified network policies. It first attempts to determine the minimum number of network segments for which segmentation is possible, and then attempts to bind objects to a minimum number of network segments required to satisfy all policies. The goal of the second task is to avoid situations where, for example, a workstation is unnecessarily added to two network segments when membership in only one of them would suffice. It is possible that the given network policies lead to conflicts. In this case, network segmentation is not possible and the policies must be changed. It is important to note that the network configuration at the present does not generate low-level technical details such as IP addresses and ports. Details such as these are expected to be set during the preparation of the cybersecurity exercise or experiment according to the concrete choice of technologies and goals.

\subsubsection{Computer installer}
\label{sec:component-copmuters}

Now that initialization of other objects has finished, and hardware requirements have been established, computers can be created. The algorithm begins by creating groups of objects called \textit{proto-computers}, which will later be split into possibly multiple computers according to hardware requirements. In each of the network segments containing server software, an initial proto-server is created, and in each network segment containing client software, proto-workstations are created for each group of objects associated with a single employee. Local software dependencies are rechecked and any missing dependencies, such as those resulting from the relaxation of the network policy \textit{RequireCollocatedLocalDependencies}, are added to proto-computers.

Once objects are grouped into proto-computers, we impose a hardware quota limit that simulates the limited capacity of computer resources in real life. If a proto-computer, when all components are installed, does not exceed the configured hardware quota limit, one computer is created and all components within the proto-computer are associated with it. Otherwise, the components are split among multiple computers.

\subsubsection{Authentication initializer}
\label{sec:component-auth}

The authentication initializer component creates credentials and their authorizations. The software templates specify whether a software package supports local or domain credentials, or whether it requires both. This is described in the \textit{data\_types} attribute. Non-root credentials are created for users to allow access to the software they use, root credentials are added for administrators to allow maintenance of the generated ITS, and additional non-root service accounts are added for server software. The rule of minimal privileges is applied, with each employee given the minimum privileges required to do their job.

If the ITS contains a domain controller (DC), which is recognizable by the \textit{ Lightweight Directory Access Protocol } (LDAP) service it provides, domain accounts are configured and stored on the DC whenever possible, and additional local accounts only when necessary. In addition, a domain admin account is created and stored on the DC that is known to the administrators. In the future, we intend to specialize the administrators and give domain admin access only to those who manage the domain. If no DC is found, local accounts are created everywhere.

\subsubsection{Security control initializer}
\label{sec:component-security}

Besides access controls, the firewall is currently the only supported security control. Therefore, the \textit{security control initializer} component creates firewall rules for the generated network, according to the following requirements:

\begin{itemize}
  \item Network services that support organizational services, e.g. Internet banking, must be able to accept connections from the Internet;
  \item If a software needs to connect to the internet, it must be able to establish connections towards the Internet; and
  \item If a client software needs network services provided by software in another network segment, a rule must be added to allow establishing such connection.
  \item If a client software needs network services provided by software in another network segment, a rule must be added to allow establishing such connection.
  \item If a DC exists, rules are added that allow software using domain credentials to establish connections to the DC.
\end{itemize}

Firewall rules are described on a high level of abstraction as a directed graph in which nodes represent software packages and edges the direction in which connections are allowed to be established. Final low-level rules can be implemented either manually or using scripts for the concrete firewall technology used in the cybersecurity exercise or experiment. Once firewall rules are created, they are grouped by endpoints and added to the ITS. Any connection not explicitly allowed by the generated rules is considered forbidden.

\subsection{Implementation}

The proposed method has been implemented as a set of Python 3.9 modules. Each of the components, corresponding to the generation phases described in Section \ref{fig:architecture}, was implemented as a Python class with a common interface that accepts templates and input parameters. This should be useful for developing a plugin system in the future. The work was done incrementally in an iterative manner, and the components were developed in the order of the corresponding phases, starting with the software chooser. As each phase was developed, the earlier phases were iteratively updated and improved.

The software chooser was implemeted using the \hl{convex optimization modelling Python package called} \textit{CVXPY} and its default \textit{ Integer Linear Programming } (ILP) optimizer \textit{cvxopt}, described in Section \ref{sec:background}. The goal is to find the \textit{approximately} cheapest combination of software that satisfies all software dependencies and all requirements imposed by employees and organizational services. At the time of writing, the problem is intentionally simplified to ignore situations such as when employees have multiple workstations. This was done for performance reasons, and we intend to include such criteria in our future work.

The ILP optimization criterion is to minimize the value $f$ in Equation (\ref{eq:ilp}). \hl{In equations throughout this paper, arrows denote vector variables, ($^\top$) is the transpose operator, and vectors are multiplied using the dot product.} Each vector in (\ref{eq:ilp}) has dimension $[n \times 1]$, where $n$ is the number of software package variants considered, with indexes corresponding to individual software package variants. Software package variants represent installations of software packages in different contexts, e.g., if the software package \textit{Python 3} can be installed on both Linux and Windows, we consider these two as different software package variants even though they represent the same software package. The variables in (\ref{eq:ilp}) are the following:

\begin{itemize}
  \item $\overrightarrow{n_0}$ is a binary vector that specifies whether software packages are installed or not;
  \item $\overrightarrow{n_n}$ specifies the total number of installations of software packages;
  \item $\overrightarrow{lc_0}$ and $\overrightarrow{lc_n}$ specify license costs for software package installations, and are further described in Appendix \ref{appendix:template-attributes};
  \item $\overrightarrow{oc_0}$ and $\overrightarrow{oc_n}$ specify operating costs for software package installations, and are further described in Appendix \ref{appendix:template-attributes};
  \item $\overrightarrow{h_q}$ specifies \textit{hardware qouta} (HQ), the abstract hardware resource requirements of software packages; and
  \item $w_h$ specifies the price \hl{of a unit amount of HQ}.
\end{itemize}

\begin{equation} \label{eq:ilp}
f 
= w_h \overrightarrow{n_n}\overrightarrow{h_q}^\top
+ \overrightarrow{n_0} (\overrightarrow{lc_0} + \overrightarrow{oc_0})^\top
+ (\overrightarrow{n_n} - \overrightarrow{n_0})(\overrightarrow{lc_n} + \overrightarrow{oc_n})^\top
\end{equation}

The variables $\overrightarrow{n_0}$ and $\overrightarrow{n_n}$ in Equation (\ref{eq:ilp}) are computed from matrix variables that describe the assignment of software to employees and organizational services using constraints. Thus, the optimization is done for the assignment, and not directly for the variables $\overrightarrow{n_0}$ and $\overrightarrow{n_n}$. Due to space limitations, technical details regarding the preparation of the ILP problem, the constraints, and the treatment of the CVXPY outputs within the software chooser are omitted. After an optimal assignment is found, final software installation records are created where necessary.

Dataset linker follows Algorithm \ref{alg:dataset-linker} to create all required dataset instances. Due to space limitations, only the high-level steps are shown. Instances are created and split multiple times according to the selected modes described in Subsection \ref{sec:rules}, with each split performed over the results of the previous split. Splits are performed either by creating initial instances of datasets if none exist, or by creating copies of instances from previous splits.


\begin{algorithm}
\caption{Dataset linkage overview}
\label{alg:dataset-linker}
\SetKwProg{funcdata}{Function \emph{instantiate\_dataset\_instances}}{}{end}
\funcdata{(employees, organizational\_services, software, Employee\_mode, Service\_mode, Software\_mode)}{
    instances = {}
    \If{supported employees exist}{
        instances = create\_employee\_instances(Employee\_mode, employees)\;
    }
    \If{supported organizational services exist} {
        \If{any(instances)}{
            instances = split\_service\_instances(Service\_mode, instances, organizational\_services)\;
        }\Else{
            instances = split\_service\_instances(Service\_mode, {dummy\_instance}, organizational\_services)\;
        }
    }
    \If{supported software exists} {
        \If{any(instances)}{
            instances = split\_software\_instances(Software\_mode, instances, software)\;
        }\Else{
            instances = split\_software\_instances(Software\_mode, \{dummy\_instance\}, software)\;
        }
    }
    \Return{instances}\;
}
\end{algorithm}

Network segmenter is implemented using the Z3 solver described in Section \ref{sec:background}. It follows Algorithm \ref{alg:network-segmenter} to assign network segments to objects generated earlier. Again, due to space limitations, only the high-level steps of the algorithm are shown. The for loop tries to find the minimum number of networks for which a solution exists. Then, a set of representative objects is selected, since it is expected that many objects have the same network segment configuration. For example, all members of an employee role subgroup (ERS) are located in common network segments, and it is sufficient to determine segments for one employee to know the rest. After segments are assigned, the algorithm uses Z3 backtracking to remove excess assignments. For example, if software A has been assigned to network segments 1 and 3, but only the assignment to segment 1 is required, this step attempts to remove the excess assignment to segment 3. Finally, the assignment for representative objects is applied to all objects. If no assignment is found until the arbitrary constant MAX is reached, network segmentation fails and input network segmentation rules must be changed.

\begin{algorithm}
\caption{Network segmentation overview}
\label{alg:network-segmenter}
\SetKwProg{funcsegment}{Function \emph{network\_segmenter}}{}{end}
\funcsegment{(objects, segmentation\_rules)}{
    \For{network\_count in 1..MAX}{
        r = get\_representative\_objects(objects)\;
        vars = create\_z3\_variables(r, network\_count)\;
        constraints = create\_z3\_constraints(r, vars, segmentation\_rules)\;
        solution = solve\_SMT\_problem(vars, constraints)\;
        groups = get\_related\_object\_groups(r)\;
        \For{group in groups}{
            solution.push()\;
            add\_join\_constraints(solution, group)\;
            solution.solve()\;
            \If{solution does not exist}{
                solution.pop()\;
            }
        }
        apply\_network\_assignment(r, solution)\;
        Transfer\_results(r, objects)\;
    }
}
\end{algorithm}

Subsequent components are implemented straightforwardly according to specifications in Section \ref{sec:architecture}. Templates are loaded from csv files, with requirements and dependencies stated as regular expressions to allow for extensibility. For example, software dependencies are stated as multiple regular expressions that match \textit{Common Platform Enumeration} (CPE) Identifires, which multiple versions of dependencies can satisfy. We chose to use JSON files to handle input parameters and outputs, as they can be manually edited in text editors and support all required data formats.

\section{Validation \& Verification}
\label{sec:validation}

During the development of the system, we continuously conducted experiments to analyze the operation of its components and rules. We tested the system with several rule configurations and found that all operations were performed as expected and that no rules were broken in the generated ITS. After that, we consulted experts to validate whether the generated ITS was a valid representation of an organizational network, and included additional rules where necessary. In this section, we describe several tests and results. 

For demonstration purposes, we describe an organization representing a fictional financial institution. It is a very simplified example, but complex enough to demonstrate the operation of the proposed system. In addition to management, the organization has an internal finance department that works with internal financial data, an external finance department that works with customers, and an IT department with administrators and developers. Two examples of employee configurations used in the tests, with 5 and 7 ERSs respectively, are shown in Table \ref{table:eval-employee-configs}.

The organization has 4 data collections describing critical data, shown in Table \ref{table:eval-dataset-configs}. ERSs in parentheses are linked only for the configuration with 7 ERS, while the rest are linked in both cases. Internal financial data (\textit{FinancialData:internal}) refers to internal financial reports and analysis and is accessible to management and the internal finance department. Banking data (\textit{FinancialData:banking}) is used for the organizational service Internet banking and is accessible to clerks, and additionally to senior developers during deployment. Emails are read by everyone, with all employees having their own Email dataset instances. Finally, the internet banking application source code (\textit{SourceCode:internet\_banking}) is accessible to developers, with each having a different version locally. The latter is configured only to demonstrate such settings for a dataset, and is usually more complex in practice.

\begin{table}
\caption{Two sample configurations of employees used in validation and performance evaluation. In evaluation, the final counts are obtained by multiplying counts from a given configuration with the multiplier specified for each experiment. Employee role subgroups are abbreviated as ERS.}
\label{table:eval-employee-configs}
\begin{center}
\begin{tabular}{ |l|l|c|c| }
\hline
\multirow{2}{*}{\begin{tabular}{@{}l@{}}Employee\\role\end{tabular}} & ERS &\multicolumn{2}{|c|}{Configuration with:} \\
& & 5 ERS & 7 ERS \\
\hline
ceo & ceo:ceo & 3 & 3 \\
ceo & ceo:financial & 2 & 2 \\
admin & admin & 2 & 2 \\
finance & finance:internal & 0 & 10 \\
finance & finance:banking &  85 & 75 \\
developer & developer:windows:senior & 8 & 4 \\
developer & developer:windows:junior & 0 & 4 \\
\hline
\multicolumn{2}{|l|}{Total employee count} & 100 & 100 \\ 
\hline
\end{tabular}
\end{center}
\end{table}

\begin{table*}
\caption{Sample configuration of datasets used in validation and performance evaluation. Abbreviations are as follows: PERS = separate instances for ERSs; PE = separate instances for employees; A = one instance for all; POS = separate instances for organizational services; PDB = separate instances for databases; PSrv = separate instances for servers.}
\label{table:eval-dataset-configs}
\begin{center}
\begin{tabular}{|l|c|l|l|l|l|l|l|l|}
\hline
    Identifier &
    \begin{tabular}{@{}l@{}}Protection\\level\end{tabular} & 
    \begin{tabular}{@{}l@{}}Datatbase\\stored\end{tabular} & 
    \begin{tabular}{@{}l@{}}Person\\mode\end{tabular} & 
    \begin{tabular}{@{}l@{}}Service\\mode\end{tabular} & 
    \begin{tabular}{@{}l@{}}Software\\mode\end{tabular} & 
    \begin{tabular}{@{}l@{}}Organizational\\services\end{tabular} &
    \begin{tabular}{@{}l@{}}ERSs\end{tabular} \\
\hline
FinancialData:internal & 5 & yes & PERS & A & PDB & & \begin{tabular}{@{}l@{}}(finance:internal)\\ceo:ceo\\ceo:financial\end{tabular}\\ \hline
FinancialData:banking & 4 & yes & A & POS & PDB & Internet\_banking & \begin{tabular}{@{}l@{}}developer:windows:senior\\ceo:financial\\finance:banking\end{tabular}\\ \hline
Emails & 2 & yes & PE & A & PSrv & & \begin{tabular}{@{}l@{}}(finance:internal)\\ceo:ceo\\(developer:windows:junior)\\developer:windows:senior\\ceo:financial\\finance:banking\\admin \end{tabular}\\ \hline
\begin{tabular}{@{}l@{}}SourceCode:\\internet\_banking\end{tabular} & 4 & no & PE & A & A & & \begin{tabular}{@{}l@{}}(developer:windows:junior)\\developer:windows:senior \end{tabular} \\
\hline
\end{tabular}
\end{center}
\end{table*}

To assess network segmentation, we described $5$ network segmentation rules. Rules were added incrementally, and each time the generated network was examined to confirm that the addition was correctly applied. The last rule (R5) was added solely for demonstration purposes. Rules were configured as follows: 

\begin{itemize}
  \item \textit{R1}: Software installed on servers mustn't be inside a network segment containing software installed on workstations;
  \item \textit{R2}: Only software providing external services, i.e. \textit{Internet banking}, is allowed to be exposed to the Internet;
  \item \textit{R3}: Data instances mustn't appear together within a network if their difference in protection levels is larger than 1;
  \item \textit{R4}: Active Directory and Exchange server must always appear together in a network segment;
  \item \textit{R5}: Network segments containing data instances with protection levels larger than 2 must not contain internet browsers.
\end{itemize}

\begin{figure}
    \centering
    \includegraphics[width=8.5cm]{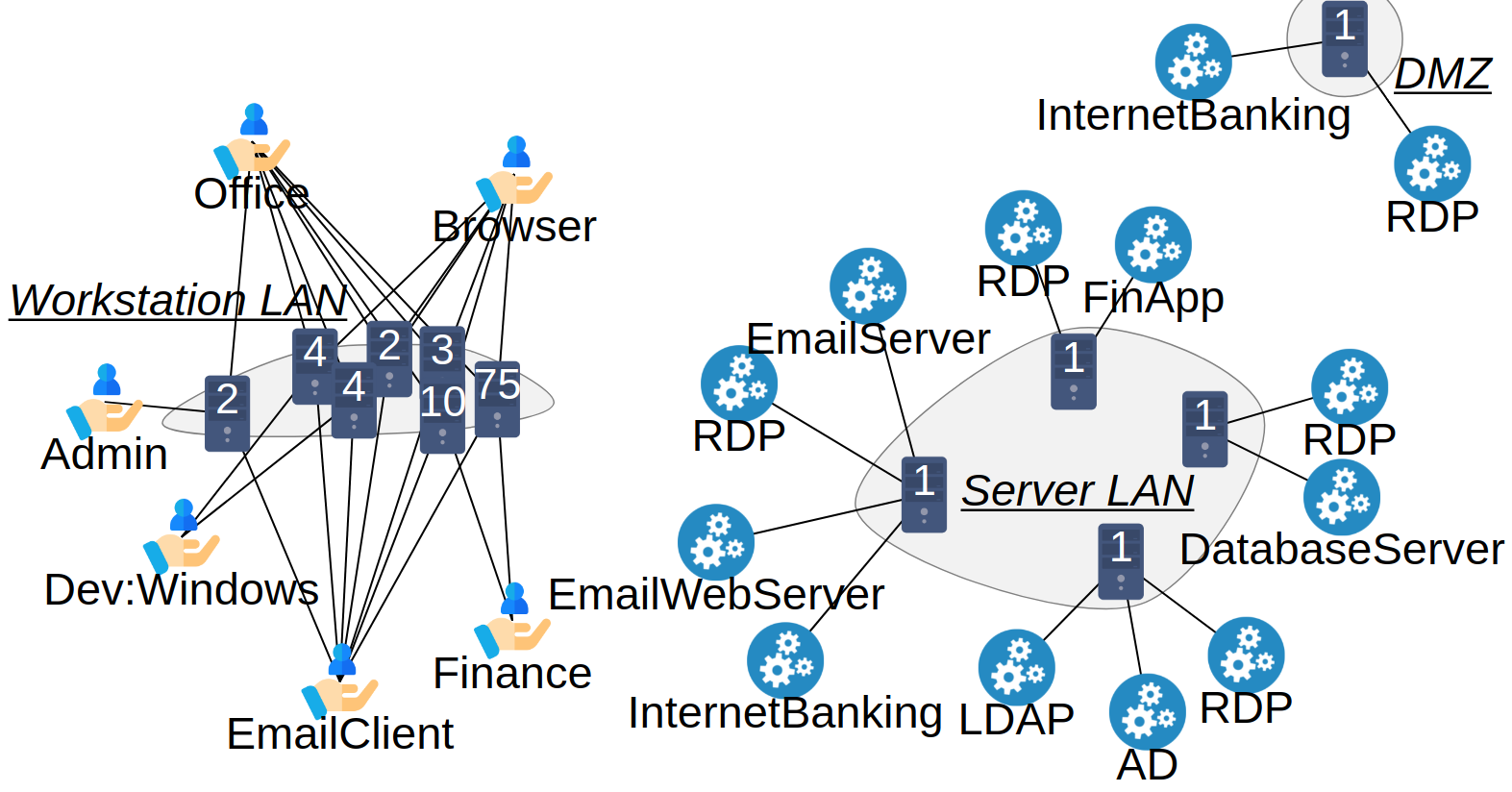}
    \caption{Output generated for 3 network rules and 7 ERS.}
    \label{fig:landscape37}
\end{figure}

\begin{figure}
    \centering
    \includegraphics[width=8.5cm]{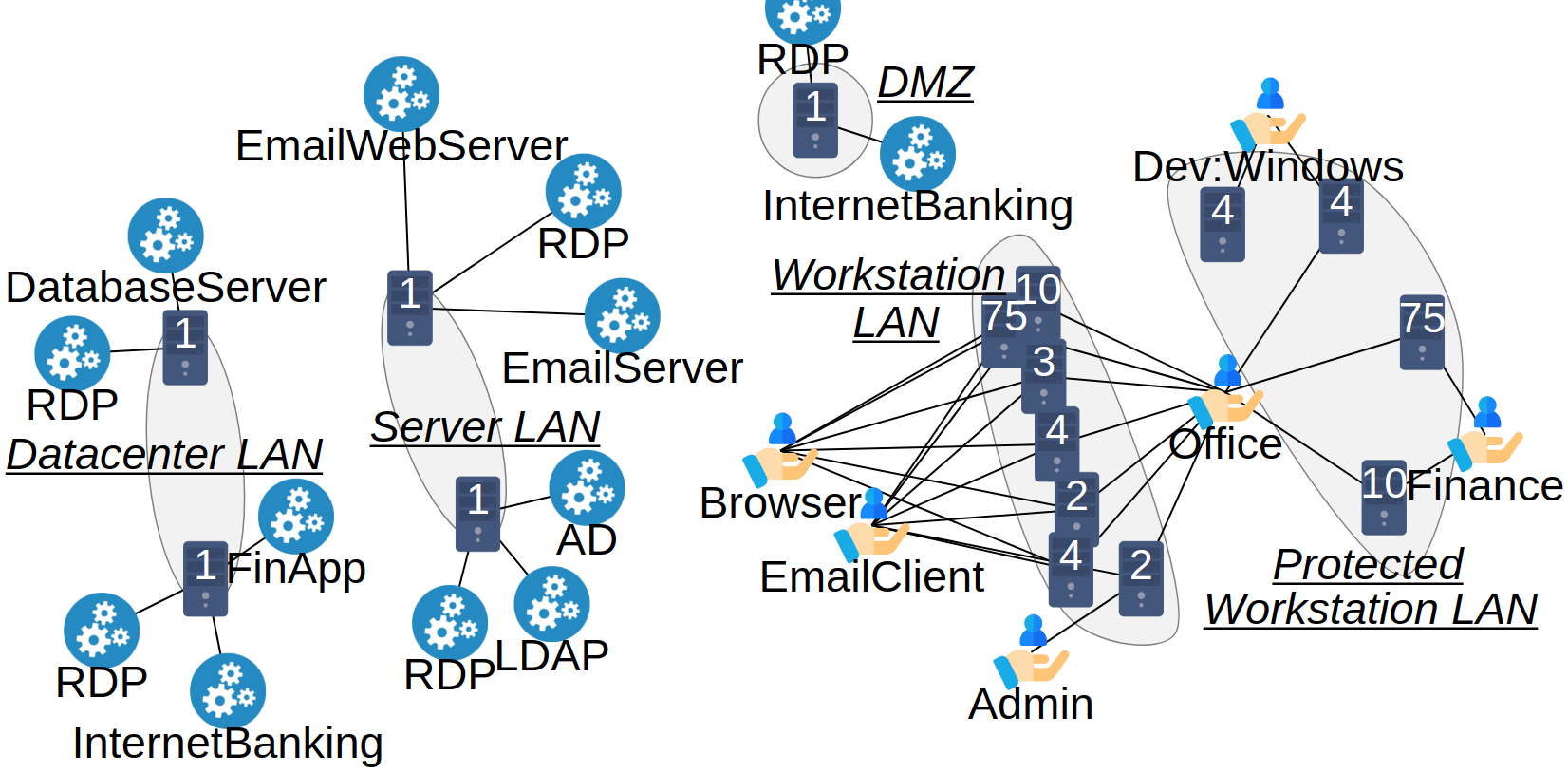}
    \caption{Output generated for 5 network rules and 7 ERS.}
    \label{fig:landscape57}
\end{figure}

Our first experiment involved network segmentation rules, with the configurations shown in Table \ref{table:eval-network-configs}. ITSs generated for the first 3 rules and for all 5 rules are shown in Figures \ref{fig:landscape37} and \ref{fig:landscape57} respectively.
PC icons represent computers, gearbox icons represent network services, hands represent user services, and enclosures with underlined italic captions represent network segments. Lines connect related objects, e.g. a line between a computer and network service signifies that that computer has server software that offers that network service. Figures are simplified to show only the high-level network architecture, with only representative computers shown. Each representative computer corresponds to a larger number of computers with similar configuration, with the count of such computers written inside the icon. For example, the number $75$ on a computer in Figure \ref{fig:landscape37} signifies that there are $75$ similar computers on which finance and browser user services are provided. Finance user services are provided by finance client applications, while browser user services are provided by web browser applications installed on those computers. Both configurations contain all 7 ERS. Three rules yield 3 network segments, while 5 rules require at least 5 network segments to satisfy the constraints. Both configurations contain one \textit{Demilitarized zone} (DMZ) network segment. The introduction of rules R4 and R5 splits the workstation LAN and server LAN segments into two parts each, as expected. The split is made so that one is allowed to contain browsers and emails, and the other is not.

\begin{figure}
    \centering
    \includegraphics[width=8.5cm]{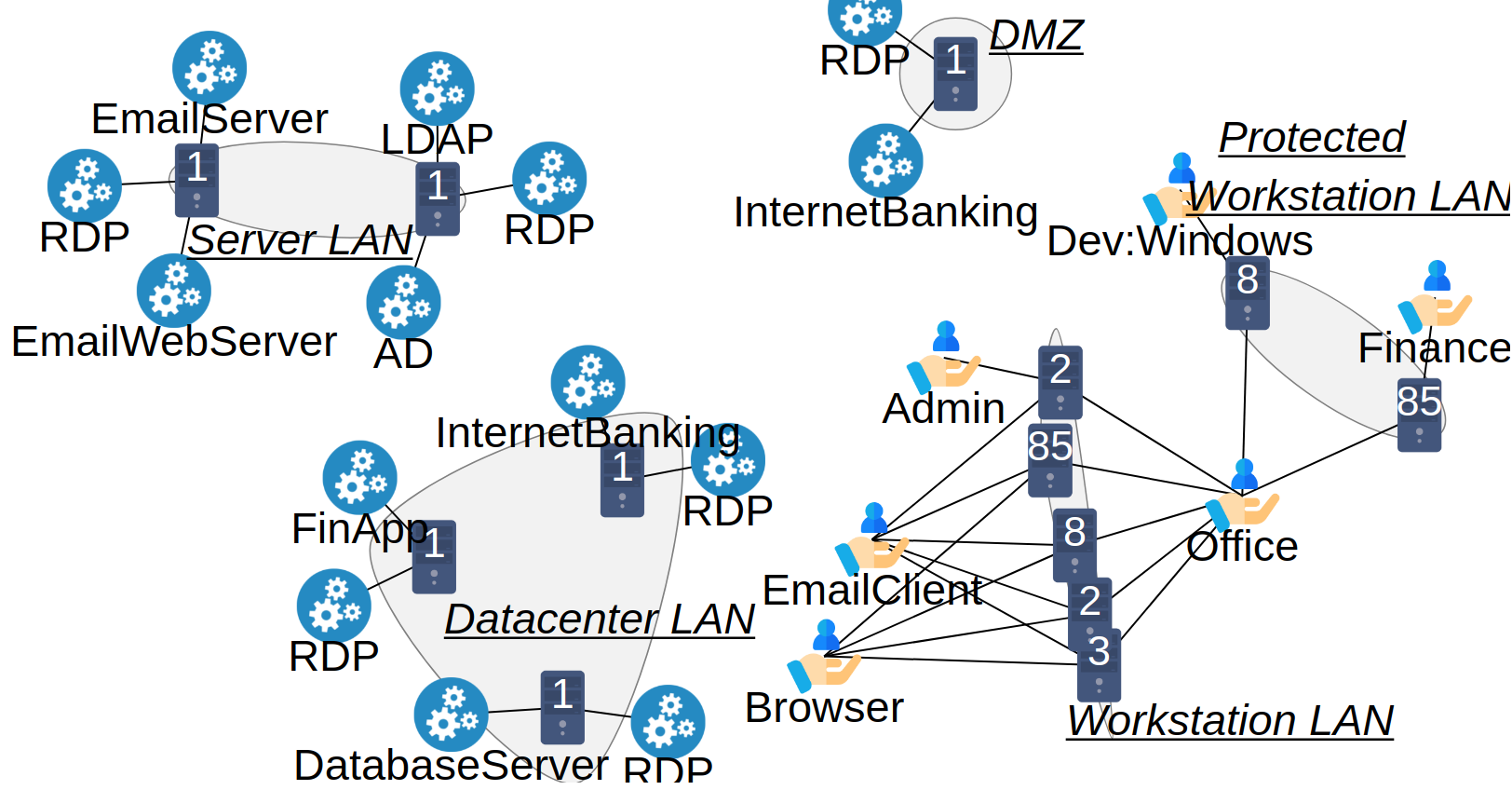}
    \caption{Output generated for 5 network rules and 5 ERS.}
    \label{fig:landscape55}
\end{figure}

The second experiment was to generate ITSs for two configurations of ERS from Table \ref{table:eval-employee-configs}. The results of applying all 5 network rules to 5 ERS and 7 ERS can be seen in Figures \ref{fig:landscape55} and \ref{fig:landscape57} respectively. 
The generated ITSs satisfy all requirements and constraints.

\begin{figure}
    \centering
    \includegraphics[width=8.5cm]{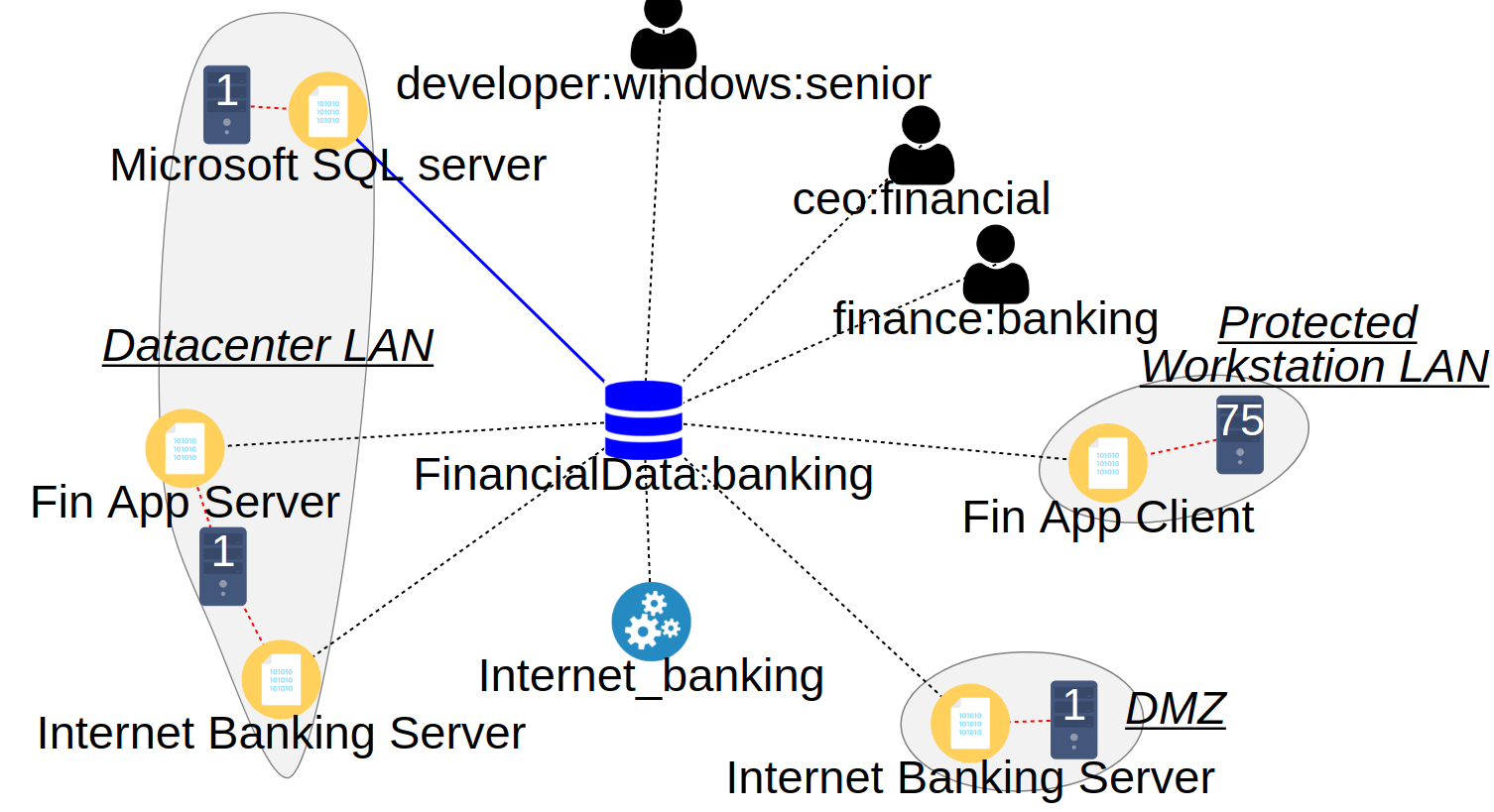}
    \caption{Example of dataset linkage.}
    \label{fig:landscape57-datasets}
\end{figure}

Dataset instance linkage for banking financial data (FinancialData:banking) is shown in Figure \ref{fig:landscape57-datasets}.
In addition to icons described earlier, database icon represents a dataset instance, user icons represent ERSs, and file icons represent software. In Figure \ref{fig:landscape57-datasets}, the gearbox icon represents an organizational service, and not a network service. Although software installations are shown only for representative computers, every computer represented by them has its own separate software installations. Only one dataset instance can be seen, corresponding to one organizational service and one database, and linked to the employees' respective client applications. The thick solid line connecting the dataset to Microsoft SQL Server denotes that the data is primarily stored there. Other dataset instances were created as expected and can be seen in Appendix \ref{appendix:landscape}.

\begin{figure}
    \centering
    \includegraphics[width=8.5cm]{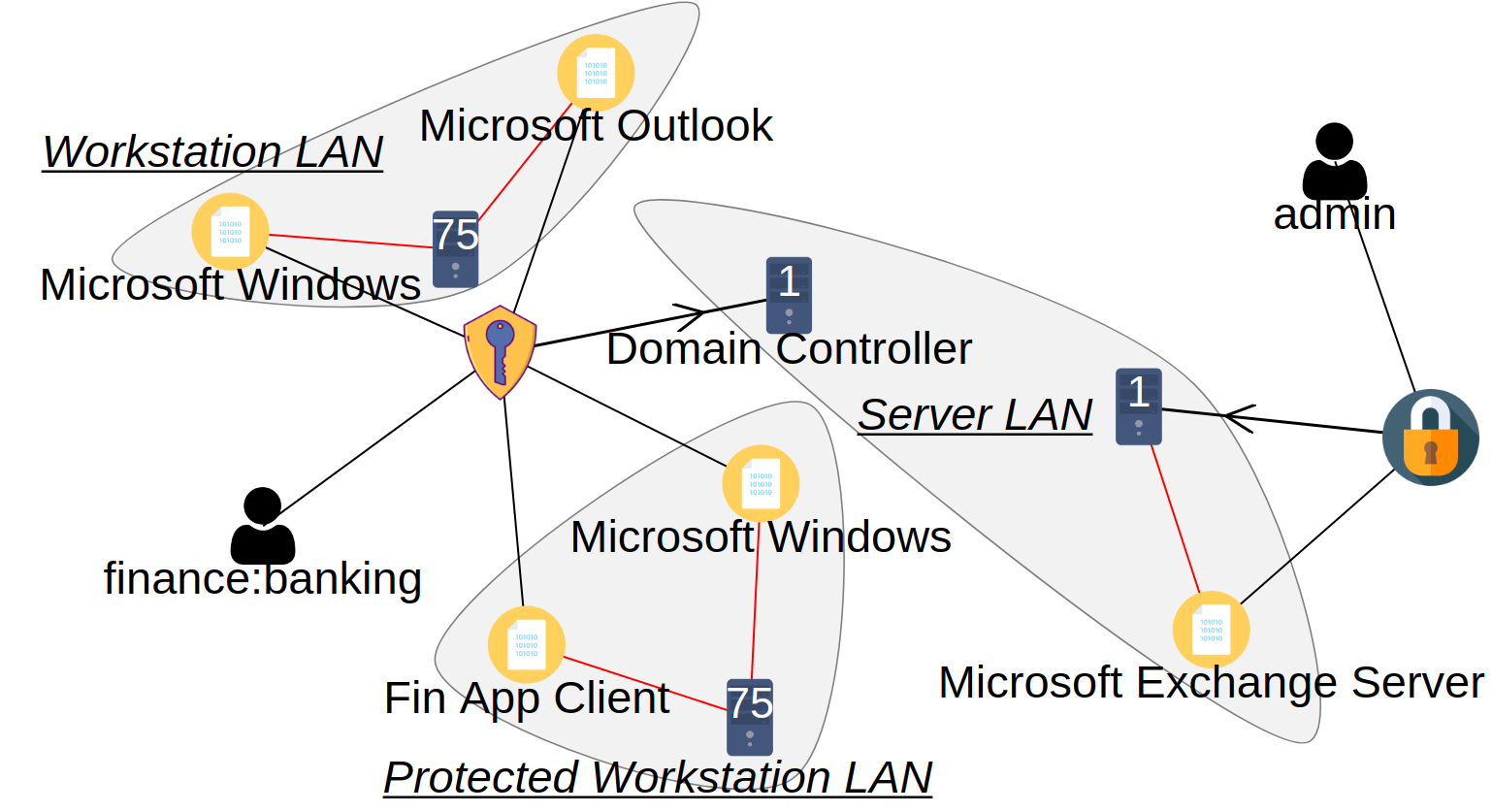}
    \caption{Example of credential linkage.}
    \label{fig:landscape57-credentials}
\end{figure}

\begin{figure}
    \centering
    \includegraphics[width=8.5cm]{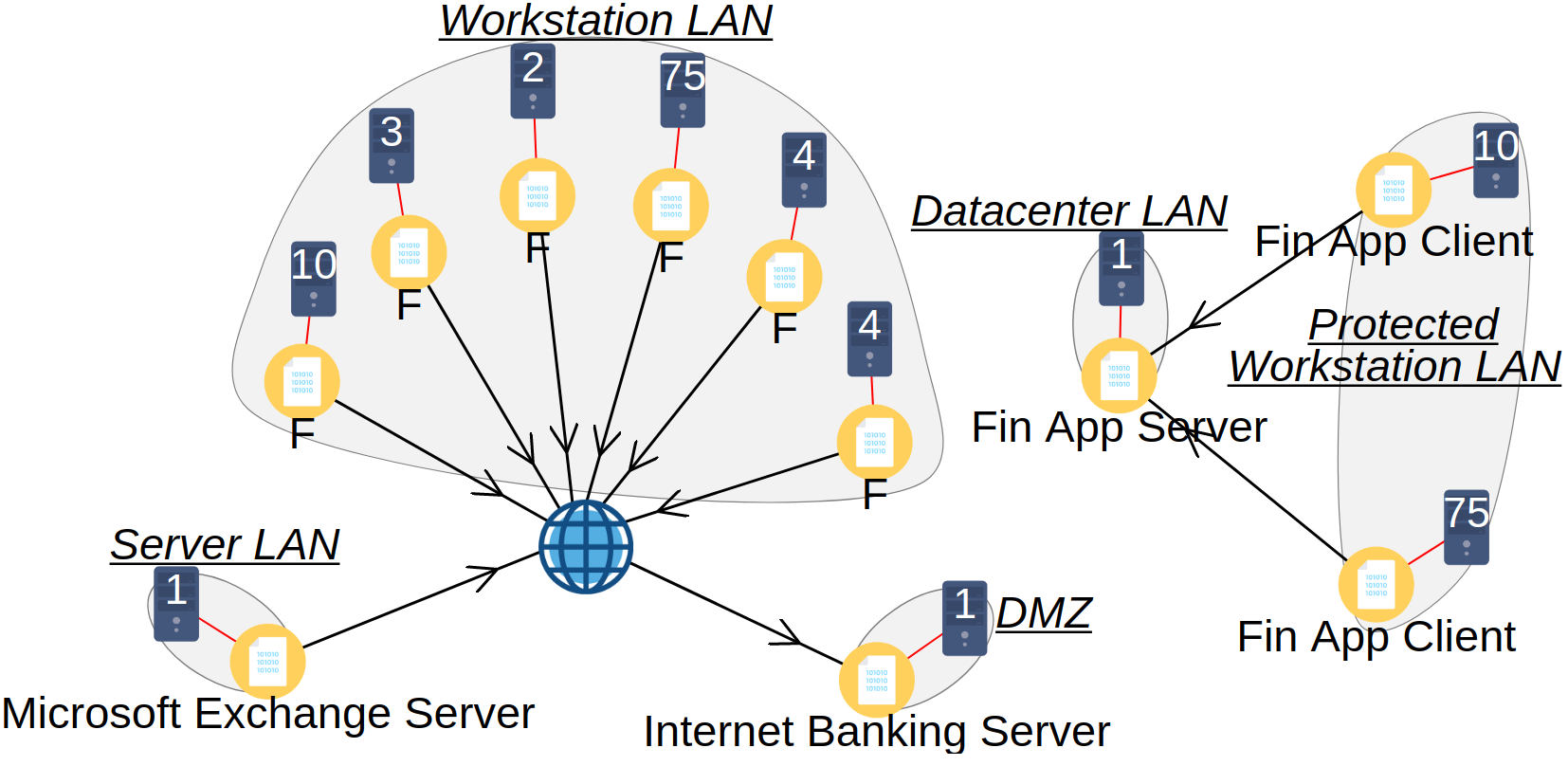}
    \caption{Examples of firewall rules. F is abbreviated for Fin app client.}
    \label{fig:landscape57-firewall}
\end{figure}

Examples of some generated credentials and firewall rules for the case with 7 ERS and 5 network segmentation rules are shown in Figures \ref{fig:landscape57-credentials} and \ref{fig:landscape57-firewall} respectively. For every operating system (OS) installation and software that explicitly requires local credentials, a local credential is created, while all software installations that support domain credentials are configured to accept credentials from the users who work on them. User credentials are represented using keyhole icons, while root credentials are represented with lock icons.
Computers on which credentials are stored are indicated with arrows, while other lines indicate software that accepts those credentials. As with computers, only representative credentials are shown. Figure \ref{fig:landscape57-credentials} shows a domain credential for a financial officer in ERS \textit{finance:banking}, stored at the DC, and linked to his/her software, alongside a local root credential for Micrsoft Exchange Server.

Figure \ref{fig:landscape57-firewall} shows firewall rules as lines with arrows indicating the direction in which connections are allowed to be established. The globe icon represents the Internet. As expected, the firewall rules created allow the Internet Banking Server to accept connections from the Internet, allow workstations with browsers to connect to the Internet, allow software clients to access their servers, and so on.
A complete visualization for the case with 7 ERS and 5 network segmentation rules, containing all types of objects, can be seen in Appendix \ref{appendix:landscape}

\section{Performance analysis}
\label{sec:performance}

Performance was evaluated by running the implemented system with a predefined range of inputs in an experimental setup. The experimental setup consisted of a computer with an Intel i7-9750H CPU, with 6 cores and 12 threads running at $2.6$ GHz, with 16 GB RAM. Since the system does not currently support multi-threaded execution, which we leave for future work, up to four instances of the system were run simultaneously, each on its own CPU core. The system and experimental scripts were implemented in Python and executed using the Python interpreter version 3.9.2.

For each experimental run, two output values were measured, duration and heap memory consumption. The duration was measured as the difference in timestamps between the moment the network generation started and the moment the results were obtained and ready to be stored. Memory consumption was measured between the two aforementioned moments using the Python package \textit{tracemalloc} and reflects only the data structures created during the generation of the ITS. This does not include the additional memory consumption by the Python interpreter and the loaded libraries, which amounts to about 200 MB additional RAM.

Input parameters were varied using a Cartesian product of two configurations of employees shown in Table \ref{table:eval-employee-configs}, two configurations of network segmentation rules shown in Table \ref{table:eval-network-configs}, and employee count multipliers as follows. A discrete numeric value, called the \textit{multiplier}, was chosen as an integer from the range $1..15$. Employee counts given in Table \ref{table:eval-employee-configs} are multiplied with the aforementioned multiplier to gain final counts for each run. The first configuration of employees includes 5 roles of employees, and the second configuration includes 7 roles of employees, both containing 100 employees in total. Hence, the total number of employees in the organization in each experiment sums up to 100 times the multiplier. Network rules from Table \ref{table:eval-network-configs} are described in Section \ref{sec:validation}.

\begin{figure}
    \centering
    \includegraphics[width=8.5cm]{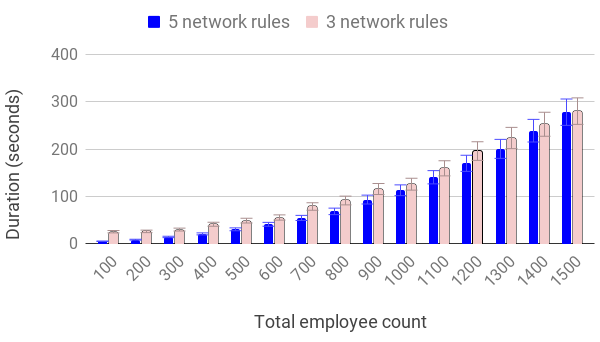}
    \caption{Average time necessary to generate topology vs employee number for five and three network rules}
    \label{fig:figure1}
\end{figure}

\begin{figure}
    \centering
    \includegraphics[width=8.5cm]{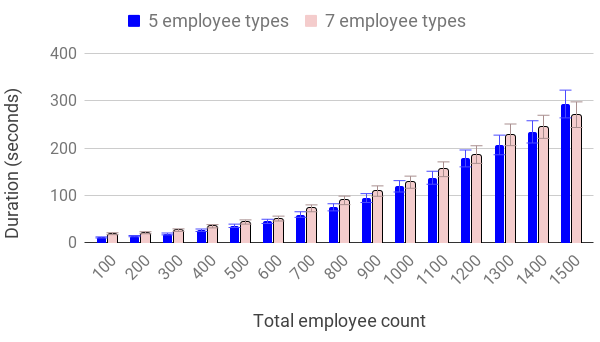}
    \caption{Average time necessary to generate topology vs employee number for 5 and 7 ERS.}
    \label{fig:figure2}
\end{figure}

\begin{figure}
    \centering
    \includegraphics[width=8.5cm]{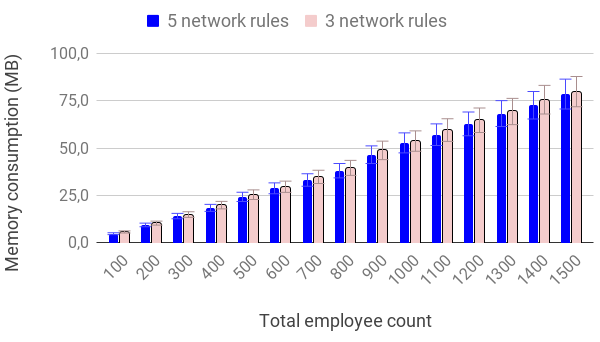}
    \caption{Average memory consumption to generate topology vs employee number for five and three network rules.}
    \label{fig:figure3}
\end{figure}

\begin{figure}
    \centering
    \includegraphics[width=8.5cm]{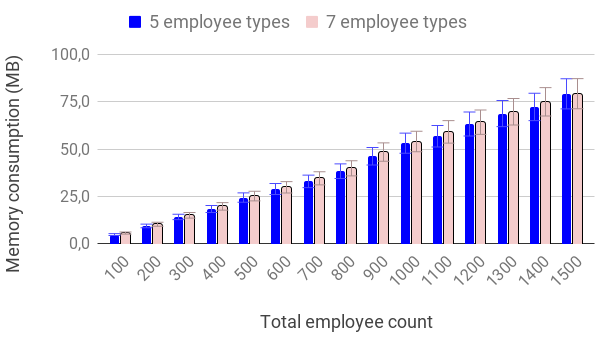}
    \caption{Average memory consumption to generate topology vs employee number for 5 and 7 ERS.}
    \label{fig:figure4}
\end{figure}

The first configuration of network rules contains $5$ rules, while the second contains $3$ rules. For each combination of input parameters, the experiment was run three times, resulting in a total of $180$ experimental runs. The resulting durations and memory consumptions can be seen in Figures \ref{fig:figure1}-\ref{fig:figure4}. All measured values, including the additional memory consumption described earlier, are well within the acceptable range for our intended use cases.

For the given range of inputs, we observe an approximately linear relationship between the total number of employees and the memory consumption, and a seemingly polynomial trend for the duration. Two configurations of employees and network rules result in small absolute differences in the measured values. Resulting ITS models and interactive visualizations of all experimental runs are available in \cite{its-dataset}. Since the dataset was created separately from the benchmark, the recorded durations and memory consumptions are expected to vary to a smaller degree.

\begin{table}
\caption{Two configurations of network segmentation rules used in the performance evaluation.}
\label{table:eval-network-configs}
\begin{center}
\begin{tabular}{|l|c|c|}
\hline
\multirow{2}{3em}{Rule} &\multicolumn{2}{c|}{Configuration with:} \\
& 5 rules & 3 rules \\
\hline
R1 & included & included \\
R2 & included & included \\
R3 & included & included \\
R4 & included & - \\
R5 &  included & - \\
\hline
\end{tabular}
\end{center}
\end{table}

\section{Discussion}
\label{sec:discussion}

The proposed system is capable of building models of ITSs with a high level of detail. In this section, we present our contributions, discuss some design decisions we made with some possible alternatives, and finally give several use cases of the proposed method and system.

Our first contribution is \textit{the proposal of the concept} of an ITS generator that generates models of ITSs with a high level of detail based on templates, rules, and input parameters. The second contribution are \textit{expert rules} for dataset instantiation and network segmentation that describe basic principles of ITS design.
Our third contribution is an \textit{implementation of an ITS generator} that not only produces measurable results, but also provides evidence that the first two contributions can lead to detailed models of ITSs. Finally, as our fourth contribution, we \textit{published a dataset} \cite{its-dataset} containing some of the generated ITS models. The published dataset will continuously be updated with newer versions of ITS models, and our goal is to include additional types of ITSs besides the aforementioned simplified financial organizations over time.

The system described in this paper is still work in progress. Presented network rules only support generating models ITSs for financial organizations, so our main goal at the moment is to expand the number of \hl{templates and} rules to support organizations from other industries. \hl{Such expansion is expected to improve generalization and show shortcomings in the current method that will need to be addressed.} Without prior research to serve as a guide, many ideas were tested while building the Generator before we reached something that works satisfactory, and we are still researching alternative methods and aspects of the Generator. As an alternative to the ILP optimizer and SMT solver, we will also consider some other algorithms, such as Constrained Clustering \cite{wagstaff2001constrained}.

\hl{
Our performance evaluation focused on organizations smaller than $1{,}500$ employees, which in 2020 constituted the vast majority of organizations within the European Union }\cite{ec2021}\hl{. Generation of ITS models for organizations with more than $1{,}500$ employees is expected to require more time and memory. Limitations in memory could be partially addressed by modelling larger organizations as a collection of smaller sub-organizations. In addition, our primary focus is on cyber ranges, which are limited in the number of computers they can simulate.
}

There are still many limitations preventing this system to create completely realistic models of ITSs. Some limitations we identified are the following ones:
\begin{itemize}
    \item Realism in context of IT systems is hard to define, so at the present we have to rely on prior experience, publicly available resources, and expert opinions to develop templates and rules.
    \item Lack of virtualization support, because of which many enterprise solutions such as cloud solutions and thin client architectures can not be generated;
    \item A small number of network rules, supporting only basic examples of organizations, as described above;
    \item Lack of support for outsourcing and client-oriented services, \hl{e.g. Virtual Private Networks (VPNs) and GitHub}.
    \item The proposed system requires all the input parameters to be specified on the input. It would be very useful if some parameters could be omitted. In that case the Generator would randomly select missing values based on some probability distributions that those parameters follow. This would allow generating many probabilistic models, all of which satisfy parameters given on the input.
    \item In the current version, models of ITSs are built from the ground up, in a single pass. Real ITSs evolve over time as organizations' they support change;
\end{itemize}

\subsection{Use cases}
\label{sec:discussion:use-cases}

We motivated this work in introduction by arguing that there are challenges with cyber ranges and validation of cybersecurity related algorithms, such as attack trees, that we could address by automation. But there are other use cases as well. In particular, we see an application in automating offense and defense using artificial intelligence and as an aid in designing new information systems or redesigning existing ones.

As we saw in Section \ref{sec:rules}, rules are the key component of the generator that embed best practices that are obeyed while designing, building and maintaining information systems in order to be as secure as possible. This makes the generator some kind of an expert system which \textit{can be used as a help tool for designers of information systems}. Specifically, when building new information systems this tool could be used to create an initial configuration suited for personal preferences of a designer. It could be also used to help improve existing information systems. For that purpose, the generator would create a target design. Then, the information system would be changed in such a way to get as close as possible to this target design.

\section{Related Work}
\label{sec:relatedwork}

In this section, we briefly address related work.
The generation of synthetic networks with desired statistical properties and realistic features resembling real networks has been an active area of research for more than two decades~\cite{calvert1997, palmer2000, medina2001, tomasik2010a, palla2010, saino2013}. In particular, Medina et al. \cite{medina2001} develop the \textit{\hl{Boston university Representative Internet Topology gEnerator}} (BRITE)~\cite{medina2001}, a topology-generation framework that, while universal, can generate networks having the hierarchical structure, node degrees, and other features similar to the topology of the Internet at the time. Tomasik et al. \cite{tomasik2010a} develop the \textit{\hl{autonomous Supelec Hierarchy Inter-domain Program}} (aSHIIP)~\cite{tomasik2010a, tomasik2010b} that focuses on faithfully modeling the relationship between the autonomous systems (AS) on the Internet. In contrast to the aforementioned works, we focus on ITSs where the network is smaller compared to the Internet. More importantly, we aim to capture complex dependencies between ITS components (rather than the statistical properties of the Internet) which is not possible with the aforementioned tools.

Works such as \cite{wagner2016,wagner2017,wagner2019,mhaskar2021} propose approaches that optimize firewall policies according to specified security and performance goals. A limitation of these approaches is that they need existing firewall policies for all resources as inputs to perform optimization. The method presented in this paper solves an orthogonal problem and does not need predefined firewall policies. It generates a collection of resources and adds only firewall rules necessary to support communication requirements inside the model. Once an ITS model is generated and its firewall policies are defined, approaches such as these can be used to further optimize network segmentation.

There are several projects that focus on generating ready-to-use systems for the purpose of education, training exercises and \textit{\hl{capture the flag}} (CTF) contests. Schreuders et al. \cite{schreuders2017} develop Security Scenario Generator~\cite{schreuders2017} that builds networks of vulnerable virtual machines based on user-defined configurations (the \emph{scenarios}). The generator supports randomization of network components, vulnerabilities and various other parameters. Alpaca~\cite{eckroth2019} uses an AI planning engine to generate \emph{vulnerability lattices} --- corresponding to sequences of steps that attacker needs to perform to achieve the goal. These and similar efforts focus on making the cyber range useful for training and education purposes. Instead, our research focuses on generating ITSs that represent plausible organizational networks and model enough details to be used for applications such as preparation of cybersecurity exercises. The other difference is that, in  the aforementioned tools, the system generation is driven by user specified configurations, while our system uses expert rules.

Finally, Russo et al. \cite{russo2020} build the \textit{\hl{Cyber Range Automated Construction Kit}} (CRACK)~\cite{russo2020} and the associated Scenario Definition Language for the purpose of designing, generating and testing complex cybersecurity scenarios. The scenario consists of a \emph{theater} roughly corresponding to our ITS model, along with injected vulnerabilities and other items relevant to the situation being modeled. Consequences of vulnerability exploits are specified by Datalog statements and the engine can check whether the scenario satisfies the desired properties --- e.g., that the attacker can reach the attack goals by exploiting a sequence of vulnerabilities present in the system. The main focus of the CRACK tool is on scenario generation and verification, assuming that the user will manually specify the theater. Hence, we view our efforts as complementary and will explore the possibility of integrating with the CRACK tool by generating ITSs that are compatible and can be used directly as CRACK theaters. 

\section{Conclusions and Future work}
\label{sec:conclusion}

In this paper, we proposed a system that receives templates and input parameters and generates a model of an organizational IT system. The generation is based on expert rules that can be selected from the existing rule collection or added manually as needed. We have successfully implemented a proof-of-concept ITS generator and demonstrated its use to generate a model of a target organizational ITS. The results were validated to ensure that all specified rules were correctly applied, and the performance of the implemented system was evaluated. The measured performance is acceptable for the intended ranges of inputs.

As future work, we intend to \hl{demonstrate the cybersecurity use case and} further test the system by developing a cybersecurity exercise theatre and scenario, validate the system with the help of domain experts, extend the proposed method and system to allow the use of probabilistic parameters, add virtualization, cloud and outsourcing support, increase the amount of technical details, and include additional network segmentation rules. Preparation of a cybersecurity exercise includes increasing the amount of technical details and choosing vulnerabilities for the scenario.  Expert assisted validation involves getting feedback and suggestions regarding the realism of the templates, rules, and generated ITS models, as well as creating scripts and configuring orchestration tools to automatically deploy and test cyber ranges based on those ITS models.
Probabilistic parameters refer to both input parameters and template attributes. Some parameters, such as optional network segmentation rules, should be selected from predefined distributions instead of being selected explicitly, and template attributes should support situations such as when some employees are given additional laptops with a certain probability. Virtualization and cloud support implies that some software packages can represent virtual computers and network components, which requires additions to several steps of the proposed method and system. Addition of technical details, such as IP addresses and logs, would make the system better suited for its use cases. Finally, the proposed set of templates and rules currently supports only financial institutions. We intend to add more rules to support a larger subset of organizations, including but not limited to banks, IT companies, and critical infrastructure operators.

As the generated models of ITSs can be used to develop and deploy cyber ranges and perform scientific experiments, our results show that ITS model generators are a promising area for future research.

\appendices

\section{Attributes of software templates}
\label{appendix:template-attributes}

As noted in Section \ref{sec:component-templates}, templates describe software packages, employee roles, and organizational services. This appendix provides a description of template attributes and examples of template objects. Software templates are described using the following attributes, with example values for software \textit{Microsoft Outlook 2019} written in brackets:

\begin{itemize}
  \item \textit{cpe\_idn}: The CPE identifier of the software package, or a unique identifier if the CPE identifier does not exist \\ \verb|("cpe:/a:microsoft:outlook:2019")|;
  \item \textit{name}: Human readable name of the software package ("Microsoft Outlook 2019");
  \item \textit{requires\_local\_software}: Information about plausible combinations of local dependencies that must be installed on a computer as prerequisites for this software package \\ ("(cpe:\textbackslash/o:microsoft:windows)(\_10|\_server\_2019).*");
  \item \textit{requires\_network\_services}: Information about variants of network services that this software must be able to access ("(EmailServer).*");
  \item \textit{provides\_network\_services}: List of network services that this software provides to other software packages and organizational services (provides none);
  \item \textit{provides\_user\_services}: List of services that this software provides to its users when installed on a workstation, distinct from network services provided for other software and organizational services ("EmailClient");
  \item \textit{requires\_hardware\_quota}: An abstract measure of the amount of hardware resources (CPU, memory, disc, etc.) required by this software package ("1");
  \item \textit{requires\_hardware\_quota\_per\_client}: Additional hardware quota required by a network service for each additional connected client ("0");
  \item \textit{lc\_0}: License cost for first installation of software ("140")
  \item \textit{lc\_n}: Additional license cost for subsequent installations of software ("140");
  \item \textit{oc\_0}: Maintenance cost for the first installation of software ("2");
  \item \textit{oc\_n}: Cost of maintenance for subsequent installations of software ("1");
  \item \textit{data\_types}: Types of data that can be processed using this software, including credential configuration ("UserAccounts:local|UserAccounts:domain", "Emails");
  \item \textit{database}: Denotes whether this software is a database management system ("0");
  \item \textit{social\_engineering\_attacks}: Denotes whether this software can be a channel for social engineering that could be used for initial access into the network ("1").
\end{itemize}

Employee roles are described using role identifiers and required variants of user services. For example, for employee role \textit{ceo}, the following service requirements are described: "(EmailCli).*", requiring an email client service, "(Office).*", requiring an Ofice suite, and "(Browser).*", requiring web browsing support. Organizational services are described in a similar manner, only with network service requirements. For example, \textit{Internet\_banking} requires network services matching "(InternetBanking).*", which ensures that an Internet banking service will be exposed to the Internet.

\section{Example of a complete model of ITS}
\label{appendix:landscape}

This appendix contains detailed visualizations of the generated ITS model. The model was generated for an organization containing $7$ employee roles and $5$ network segmentation rules, as described in Section \ref{sec:validation}. Only representative computers are shown, because there are in total $100$ employees and $199$ computers, each with distinct software installations, and in some cases even distinct dataset instances.

\begin{figure*}
    \centering
    \includegraphics[angle=90,origin=c,height=16cm]{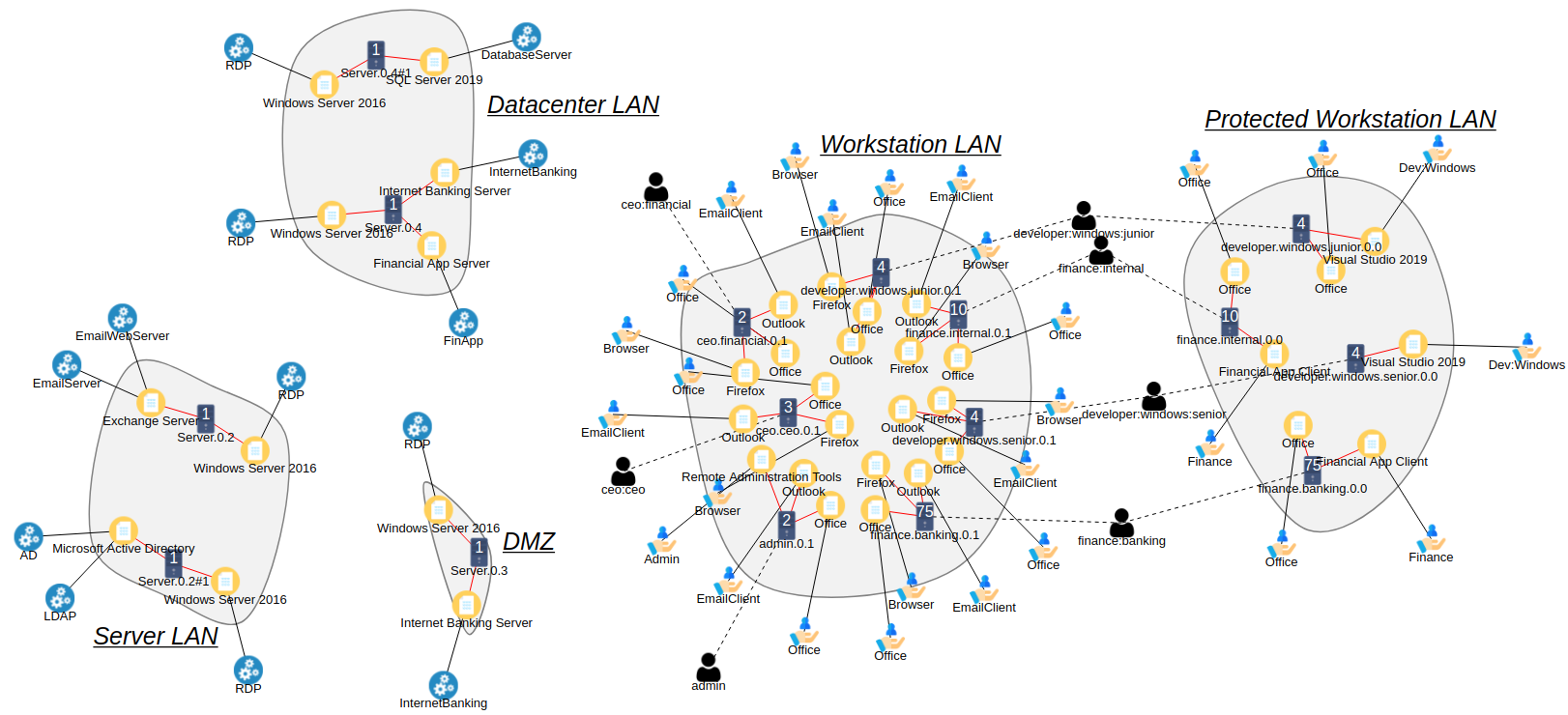}
    \caption{Generated computers and software installations with linkage to ERS, user services, and network services. Organizational service \textit{Internet banking} linked to the \textit{Internet Banking Server} software in the DMZ is not shown.}
    \label{fig:appendix-landscape}
\end{figure*}

\begin{figure*}
    \centering
    \includegraphics[angle=90,origin=c,height=17.5cm]{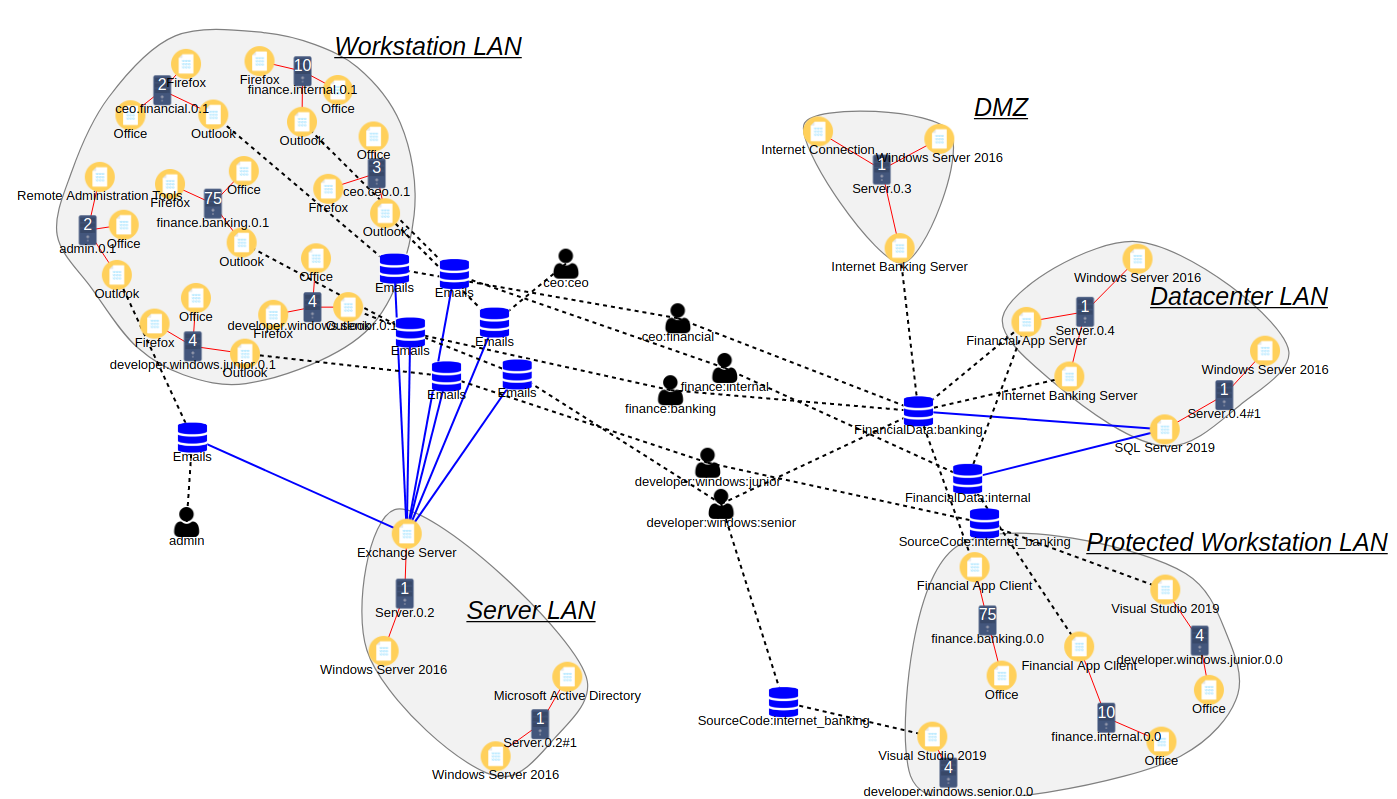}
    \caption{Generated dataset instances with linkage to ERS, organizational services, and software installations.}
    \label{fig:appendix-datasets}
\end{figure*}

\begin{figure*}
    \centering
    \includegraphics[angle=90,origin=c,height=18cm]{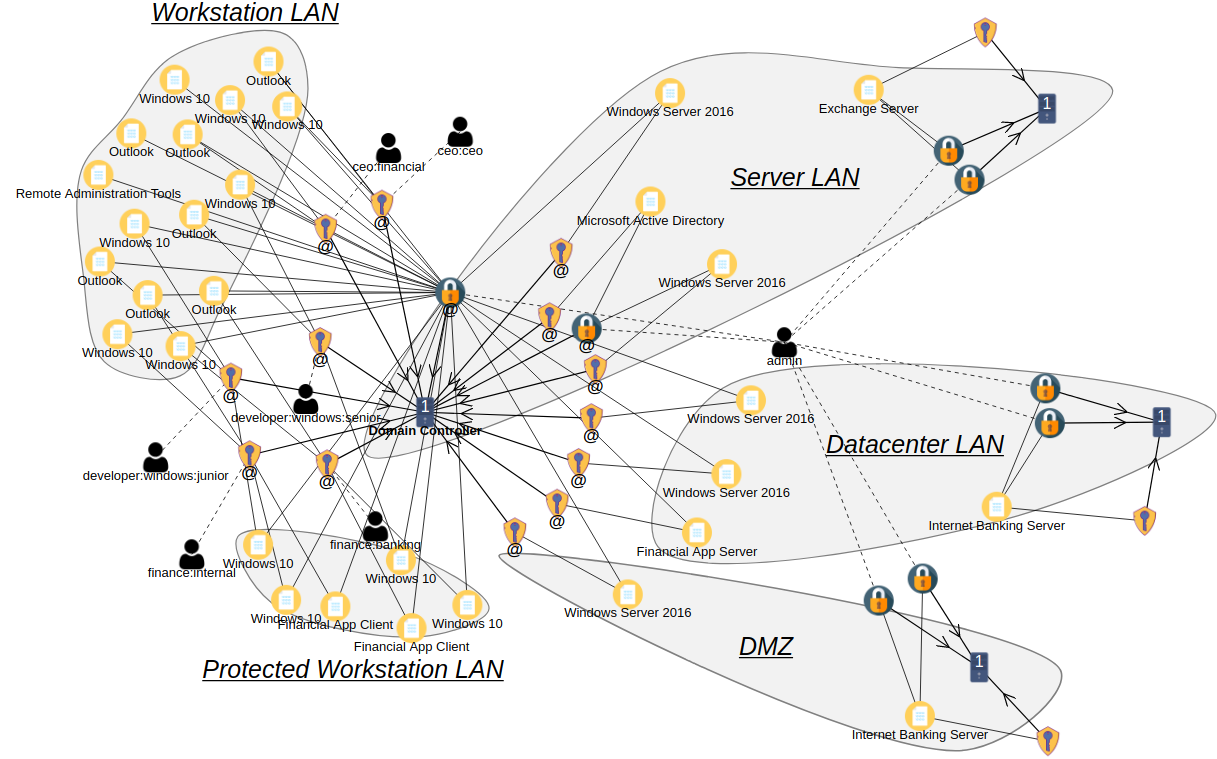}
    \caption{Generated credentials with linkage to ERS, software installations, and computers on which they are stored.}
    \label{fig:appendix-credentials}
\end{figure*}

\begin{figure*}
    \centering
    \includegraphics[angle=90,origin=c,height=17.5cm]{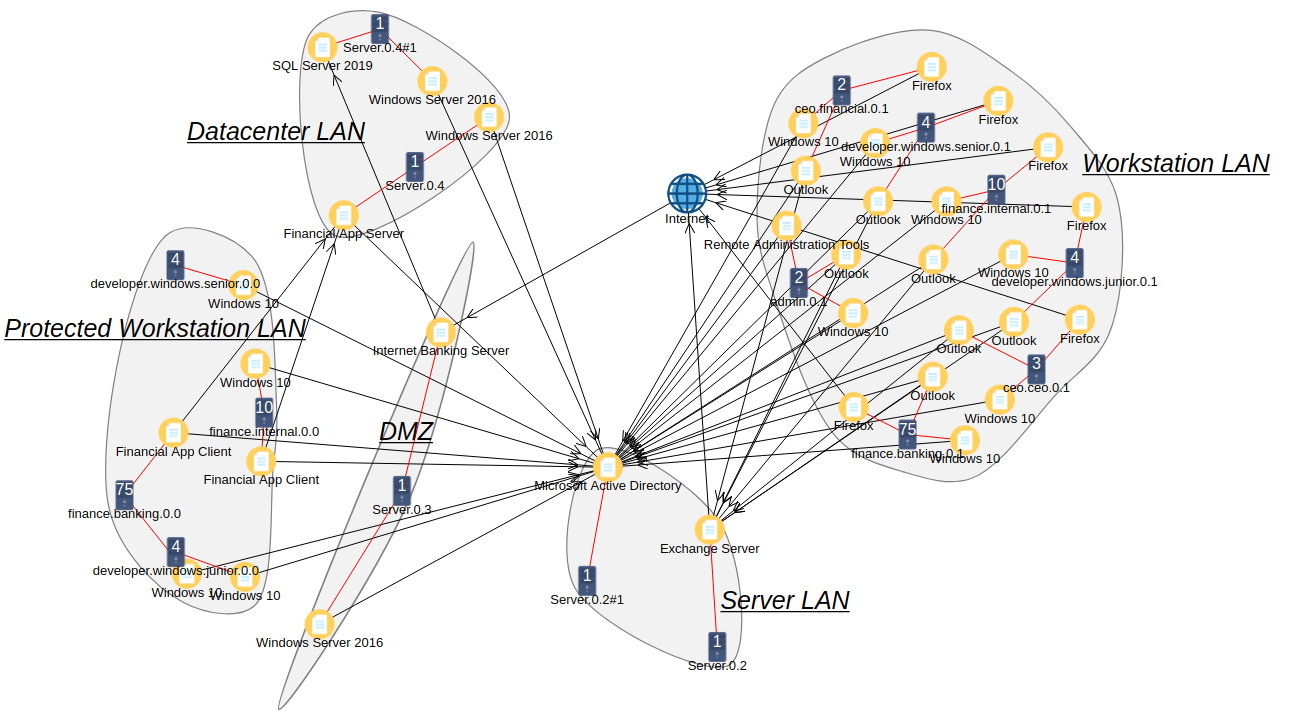}
    \caption{Generated firewall rules, represented by lines with arrows. Arrows denote the direction in which connections are allowed to be established. Rules for administration are omitted for clarity.}
    \label{fig:appendix-firewall}
\end{figure*}

Figure \ref{fig:appendix-landscape} provides an overview of computers and software installations, 
Figure \ref{fig:appendix-datasets} shows dataset instances with linkage, Figure \ref{fig:appendix-credentials} shows generated credentials, and Figure \ref{fig:appendix-firewall} shows generated firewall rules. Icons and signs are the following:

\begin{itemize}
    \item ERS are represented by user icons. ERS can contain a large number of employees that are not shown, with every employee having distinct computers, credentials, etc.
    \item Organizational services are not shown. The only organizational service is Internet Banking, and it is linked to the DMZ Internet banking server and the instance of \textit{FinancialData:banking} dataset linked to it.
    \item Network segments are represented by encircled shaded areas with names underlined and written in italic.
    \item Representative computers are denoted using computer icons, with numbers inside them signifying the count of similar computers represented by the representative one.
    \item Software installations are represented using yellow file icons, and linked to computers where they are installed.
    \item Network services are represented with gearbox icons, and linked to software installations that offer them.
    \item Dataset instances are represented using database icons, with a thick connector line indicating the location where they are primarily stored, and dashed lines indicating locations on which they can be accessed and people (or ERS) who can access them.
    \item Ordinary credentials are shown as keyhole icons and linked using ordinary lines to software packages that accept them and employees who use them, and linked with a directed line to the computer they are stored at.
    \item Privileged credentials (i.e. root) are shown as lock icons, with the same pattern of linkage as ordinary credentials.
    \item Domain credentials are indicated using the "@" sign.
    \item Firewall rules are shown as lines with arrows that signify the direction in which establishing connections is allowed. Firewall rules do not need to be defined between objects within the same network segment, because network traffic filtering is performed only when objects from different network segments are involved.
\end{itemize}


\section*{Acknowledgment}

The authors would like to thank Bruno Bijelić for his work on the visualizations in this article. Icons in diagrams were obtained from \href{https://www.flaticon.com}{www.flaticon.com}. Attribution lines required by license terms are the following:

\begin{itemize}
    \item Icon made by \href{https://www.flaticon.com/authors/freepik}{Freepik} from \href{https://www.flaticon.com}{www.flaticon.com}
    \item Icon made by \href{https://www.flaticon.com/authors/flat-icons}{Flat Icons} from \href{https://www.flaticon.com}{www.flaticon.com}
    \item Icon made by \href{https://www.flaticon.com/authors/vectors-market}{Vectors Market} from \href{https://www.flaticon.com}{www.flaticon.com}
    \item Icon made by \href{https://www.flaticon.com/authors/dinosoftlabs}{DinosoftLabs} from \href{https://www.flaticon.com}{www.flaticon.com}
\end{itemize}

This work was supported in part by 
the research and development project Cyber Conflict Simulator, co-financed by the EU under Grant KK.01.2.1.01.0054.

\bibliographystyle{IEEEtran}
\bibliography{bibliography}

\begin{IEEEbiography}[{\includegraphics[width=1in,height=1.25in,clip,keepaspectratio]{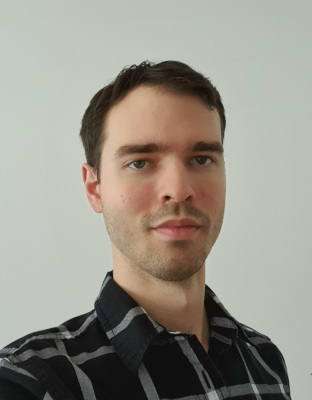}}]{IVAN KOVAČEVIĆ} was born in Rijeka, Croatia. He received his B.S. and M.S. degrees from University of Zagreb, Faculty of Electrical Engineering and Computing in 2015 and 2017 respectively, and currently works there as a Research Associate. From January 2018 to December 2020, he worked on the research project CCS (Cyber Conflict Simulator), aiming to simulate cyber incidents for cyber security exercises. At the present, he works on two projects. The first is related to security of critical infrastructures, and the second aims to find compromised websites within the Croatian cyberspace. His scientific and professional interests lie in the fields of cybersecurity, computer networks, and software engineering, with his doctoral research focused on automating preparation of cybersecurity exercises. 
\end{IEEEbiography}

\begin{IEEEbiography}[{\includegraphics[width=1in,height=1.25in,clip,keepaspectratio]{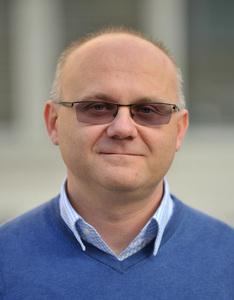}}]{STJEPAN GROS} is an assistant professor at the Faculty of Electrical Engineering and Computing, University of Zagreb. His scientific and professional interests are in the field of information and cyber security and the application of advanced methods in solving problems in these areas. He also deals with issues of research and development management. Stjepan Groš participates in the implementation of several EU-funded projects in the field of cyber security, focusing on the study of attacker behavior and automation using machine learning algorithms. He has published a number of papers in the field of information security, computer networks and operating systems.
\end{IEEEbiography}

\begin{IEEEbiography}[{\includegraphics[width=1in,height=1.25in,clip,keepaspectratio]{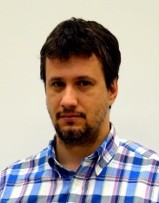}}]{ANTE ĐEREK} is an assistant professor at the Faculty of Electrical Engineering and Computing, University of Zagreb. His research focuses on applying formal methods to problems in computer security, privacy and cryptography. He participates in a number of national and EU-funded projects in the area of computer security.
\end{IEEEbiography}

\EOD 

\end{document}